\documentclass[prb,aps,twocolumn]{revtex4}
\usepackage{tabularx}
\usepackage{bm}
\usepackage{euscript}
\usepackage{epsfig,psfrag,subfigure}
\usepackage{graphicx}
\usepackage{color}
\usepackage{amsfonts}
\usepackage{exscale}
\usepackage{amsbsy}
\usepackage{stmaryrd}
\usepackage{amsmath}

\def\be{\begin{equation}}       \def\ee{\end{equation}}
\def\bea{\begin{eqnarray}}      \def\eea{\end{eqnarray}}
\def\ba{\begin{array} }
\def\ea{\end{array} }
\def\bnum{\begin{enumerate} }
\def\enum{\end{enumerate}}

\def\=>{\Rightarrow}
\def\>{\rightarrow}

\def\eye2{Fathbb{I}}

\begin{document}

\title{\bf 
Enhancement of superconductivity near a nematic quantum critical point}

\author{S. Lederer$^1$, Y. Schattner$^2$, E. Berg$^2$, and S. A. Kivelson$^1$}

\affiliation{$^1$Department of Physics, Stanford University, Stanford, California 94305, USA}

\affiliation{$^2$Department of Condensed Matter Physics, Weizmann Institute of Science, Rehovot, Israel 76100}

\begin{abstract}
We consider a low $T_c$  metallic superconductor weakly coupled to the soft fluctuations associated with proximity to a nematic quantum critical point (NQCP).  We show that: 
1) a BCS-Eliashberg treatment remains valid outside of a parametrically narrow interval about the NQCP; 2) the symmetry of the superconducting state (d-wave, s-wave, p-wave) is typically determined by the non-critical interactions, but $T_c$ is enhanced  by the nematic fluctuations in all channels;  3) in 2D, this enhancement grows upon approach to criticality up to the point at which the weak coupling approach breaks-down, but in 3D the enhancement is much weaker. 
\end{abstract}
\date{\today }
\maketitle

In both the hole-doped cuprate\cite{howald-2003,xia-2007,hinkov-2007,taillefer-nematic-2009,lawler-2010,wu-2014,grissonnanche2014,ramshaw2014,fujita2014} 
and 
Fe-based\cite{fisher-2010,chu2012,yoshizawa2012,matsuda2013} high temperature superconductors, there is
   evidence
  of a nematic quantum critical point (associated with the breaking of point group symmetry) at a critical doping, $x_c$,  which is close to the ``optimal doping'' 
  at which the  superconducting $T_c$ is maximal.  These materials are complicated, strongly coupled systems with many intertwined ordering tendencies\cite{berg-2009b,Fradkin-2012,davis2013,palee2014}, and in which quenched disorder plays a 
  role in 
  some aspects of the physics\cite{nie-2013,wu-2014}.

Thus motivated by experiments, but without  pretense that the theory is {\it directly} applicable to these materials, we study the situation in which a low $T_c$ metallic superconductor is weakly coupled, with coupling constant $\alpha$, 
to collective modes representing  the soft fluctuations of a system in the neighborhood of a nematic quantum critical point (NQCP).  
Here, the effective interaction in the Cooper channel consists of the sum of a non-retarded, non-critical piece $V^{(0)}$, and a critical piece, $V^{(ind)}$, which is increasingly peaked at small momentum and energy transfer the closer one approaches to the NQCP.  
 The peak width 
 as a function of wave-number and frequency is, respectively, $\kappa\equiv \xi^{-1}$ and  $\Omega\sim \xi^{-z}$, where $\xi$ is the nematic correlation length, 
 $z$ is the dynamical critical exponent, and where on the ordered side of the NQCP the nematic transition temperature is comparable to $\Omega$.
  For small $\alpha$,
outside of a parametrically narrow regime about criticality, 
 the induced interactions among the electrons 
 can be computed without needing to worry about the feedback effect of the fermions on the collective modes. 
 
 We thus gain analytic control of the problem in a parametrically broad quantum critical regime, though not in a small window of {\it metallic} quantum criticality, (see Fig. 1). 
 In 
the regime of control, 
 $V^{(ind)}$ is weak ($T_c \ll \Omega$) and so can be treated in the context of BCS-Eliashberg theory, or equivalently,\cite{shankar,polchinski} perturbative renormalization group (RG).  
 The nematic modes 
 play a role similar to that of
 phonons in a conventional superconductor, with the 
   difference that $V^{(ind)}$ is strongly $k$ dependent in such a way that it is attractive in all pairing channels, and so enhances $T_c$ in whatever channel is favored by the non-critical interactions. 
 The enhancement grows rapidly upon approach to criticality in 2D, and somewhat more slowly in 3D. 
  \begin{figure}%
\centering
\includegraphics[width=4.3cm]{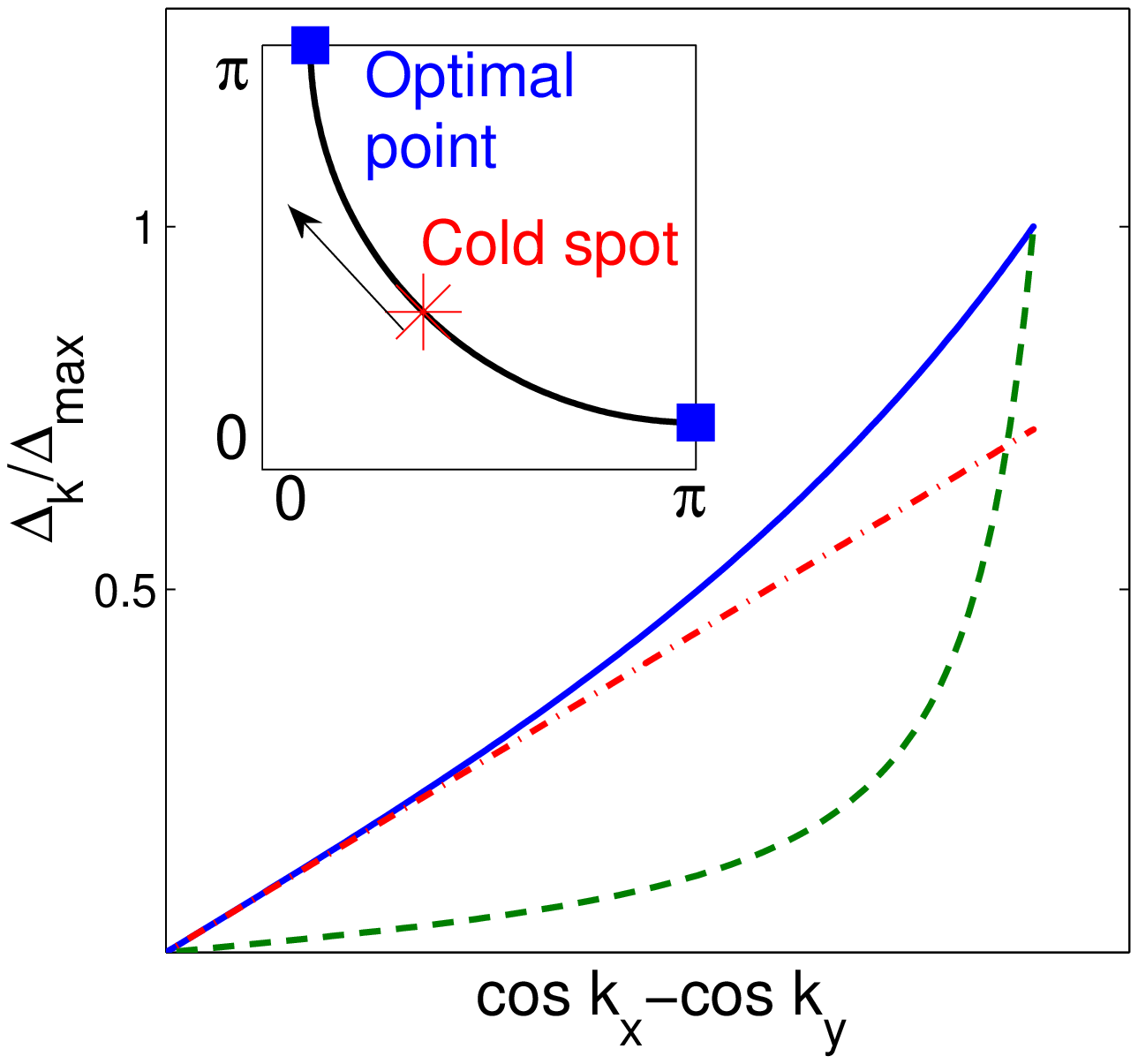}%
\raisebox{5pt}{\includegraphics[width=3.7cm,height=3.5cm]{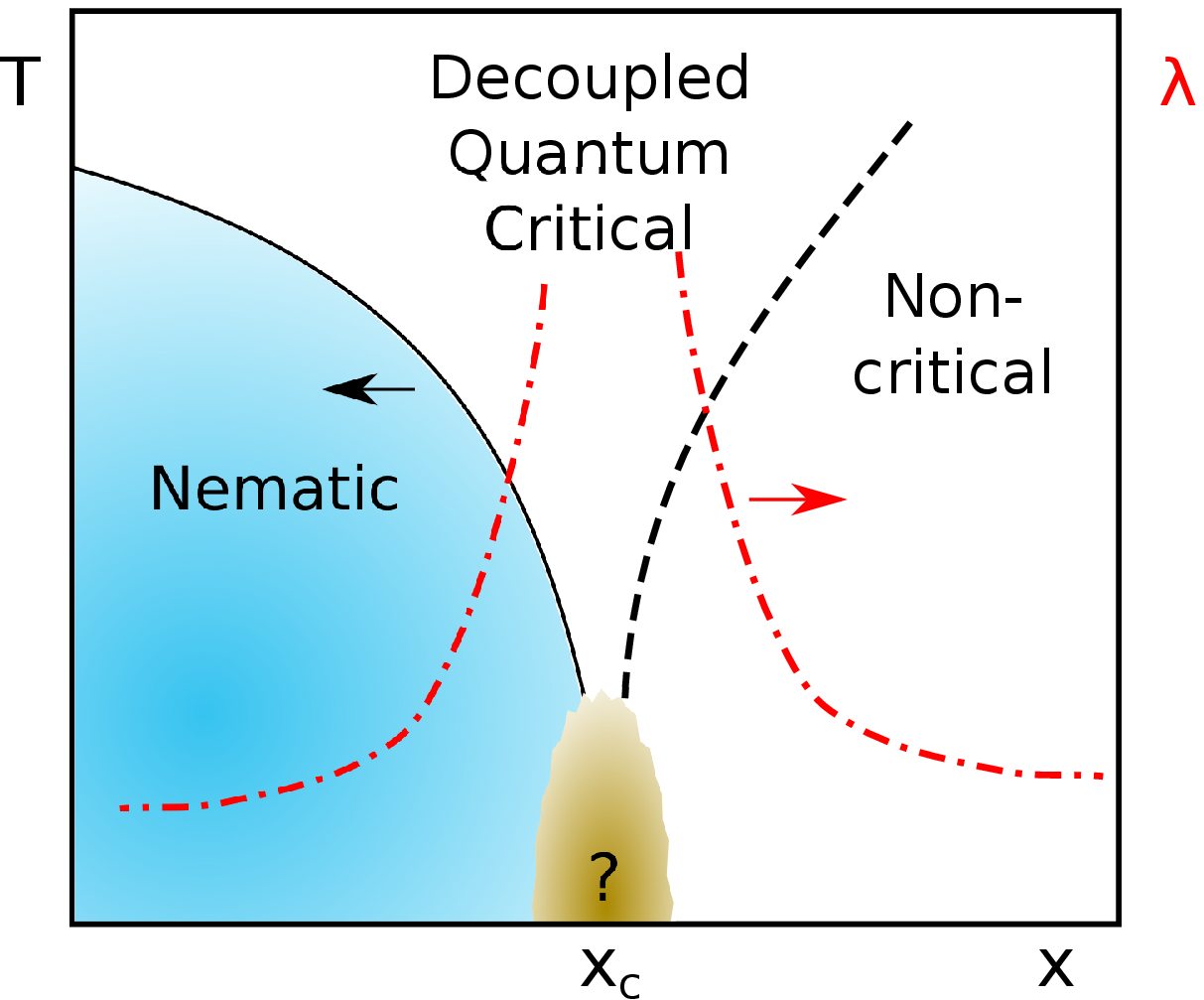}}
\caption{{\bf a)} Gap function vs. $\cos k_x - \cos k_y$ along the Fermi surface, obtained by solving Eq.~\ref{eigenstates} in 
2D. The non-critical interaction 
$\Gamma^{(0)}$ is taken to 
favor a d-wave gap with form factor $\cos k_x - \cos k_y$ (dot-dashed line) and have a strength such that  $\lambda^{(0)} =10\alpha^2\approx0.05$. 
The solid and dashed lines
are, respectively, for $k_F\xi=10$ (``weak enhancement'') and $k_F \xi =100$ (``strong enhancement''). 
{\bf Inset: } Cartoon of the Fermi surface
of the cuprates; the star and squares indicate, respectively, 
a ``cold spot" of the nematic $d_{x^2-y^2}$ form factor where 
 $\Gamma^{(ind)}$ 
 vanishes by symmetry, 
and 
 ``optimal points" $\hat k_{opt}$ where $\Gamma^{(ind)}$ is strongest. 
{\bf b)} Schematic phase diagram. The solid line shows the nematic transition temperature. Outside of the central shaded region, the quantum critical regime can be described in terms of a Wilson-Fisher fixed point weakly coupled to a Fermi liquid. Within the shaded region, this picture breaks down, and a different description (possibly in terms of a different, strongly coupled fixed point) is needed. $T_c$ is much smaller than any temperature scales pictured, but varies dramatically. The dashed-dotted line shows the behavior of $\lambda$, the pairing eigenvalue (Eq.~\ref{lambda}). 
}

\label{fig:cont}%
\end{figure}

{\bf The Model:} We consider a system described by the Euclidean effective action
\be
S[\phi,\bar\psi,\psi] = S_{el}[\bar\psi,\psi]+S_{nem}[\phi] + S_{int}[\phi,\bar\psi,\psi]
\ee
where 
$ S_{el}$ is the action of itinerant electrons with 
an assumed weak interaction in the Cooper channel, $V^{(0)}(\vec k,\vec k^\prime)$, $S_{nem}$  is the action of the nearly critical nematic mode $\phi$, 
and
\be
S_{int}=\alpha\int d\tau\frac{d\vec k}{(2\pi)^d}\frac{ d\vec q}{(2\pi)^{d}}f(\vec q, \vec k)\phi_{\vec q} \bar \psi_{\vec k+\vec q/2}\psi_{\vec k-\vec q/2}
\ee
where we have suppressed the spin index on the fermion fields.  We consider Ising nematic order of $d_{x^2-y^2}$ symmetry in a system with tetragonal symmetry, which implies that $f(\vec q,\vec k)$ is  
 odd under rotation by $\pi/2$ and under reflection through  $(1,\pm 1,0)$ mirror planes, but even under inversion, time-reversal, and reflection through  $(1,0,0)$ and $(0,1,0)$ mirror planes. 
  Since the physics near criticality is dominated by long-wave-length nematic fluctuations, the coupling constant can 
   be replaced by its value at $|\vec q|\to 0$, $f(\vec 0,\vec k)\equiv f(\vec k) \sim [\cos(k_x)-\cos(k_y)]$.  
   (Note: this form factor reflects the symmetry of the nematic order, and is unrelated to the symmetry of the pair wavefunction). 

This effective action already represents a coarse-grained version of the microscopic physics.  In particular, since the nematic phase 
breaks
the point-group symmetry of the crystal, $\phi$ generally involves  collective motion of both the electron fluid and the lattice degrees of freedom, with relative weights that depend on microscopic details.
In the absence of coupling to low energy electronic degrees of freedom ($\alpha=0$), we suppose that as a function of an externally controlled parameter $x$ (which could be doping concentration, pressure etc.), there is a quantum phase transition from a nematic phase for $x<x_c$ in which $\phi$ is condensed, to an isotropic phase for $x>x_c$.  Thus, the dynamics of $\phi$ 
are characterized by $d+1$ dimensional Ising exponents with 
 $z=1$.

In the fermion sector we 
introduce a cutoff,  $W$,
defined as the energy scale of the non-critical portion of the electron-electron interaction, $V^{(0)}$.  
For instance, if $V^{(0)}$ is mediated by short-range spin-fluctuations,\cite{keimer} the cutoff energy is 
proportional to the
exchange coupling $J$. 
 Restricting fermion energies to lie below $W$ justifies both neglecting all irrelevant couplings (other than those in the Cooper channel) and treating the remaining interactions as non-retarded.  
(
More generally, we should include Fermi liquid parameters in $S_{el}$, but we will neglect these for simplicity.) 

{\bf Effective Interactions:}  In the small $\alpha$ limit, beyond a parametrically narrow interval about criticality, 
 one can integrate out the nematic modes perturbatively to produce an  effective action for the electrons alone.  In the disordered phase ($x > x_c$), the leading order effect is an additive four-fermion term proportional to $\alpha^2\chi(\vec q,\omega)$, which in the Cooper channel 
results in the net interaction
\bea
V\left( \vec k+\frac{\vec q}{2},\vec k-\frac{\vec q}{2},\omega\right) = & V^{(0)}\left( \vec k+\frac{\vec q}{2},\vec k-\frac{\vec q}{2}\right)\\
 &-\frac{1}{4} \alpha^2 |f(\vec q,\vec k)|^2\chi(\vec q,\omega) \nonumber
\eea
where $\chi(\vec q, \omega)$ is the nematic susceptibility, which is peaked at $\vec q=\vec 0$ and $\omega=0$.  With the usual definition of the critical exponents, 
 $\chi(\vec 0,0) \equiv \chi_0$ diverges as $\delta x \equiv (x-x_c) \to 0$ as $\chi_0 \sim |\delta x|^{-\gamma}$, and falls as a function of increasing $|\vec q|$ and $|\omega|$ as
$\chi(\vec q,\omega) \sim \chi_0 \ [\Omega^2/(c^2q^2+\omega^2+\Omega^2)]^{1-\eta/2}$. 
 The $\vec q$-space width of 
 $\chi(\vec q,0)$ is thus $\kappa =  \xi^{-1} \sim \Omega^{1/z} \sim |\delta x|^{\nu }$. (From scaling, $1-\eta/2=\gamma/2\nu$.) The Ising critical exponents are $\{\nu,\eta,\gamma\}=\{1/2,0,1\}$ for $d=3$, and $\{\nu,\eta,\gamma\}\approx\{0.63,0.03,1.23\}$ for $d=2$.\cite{Privman1991}

There are also
other (mostly irrelevant) four-fermion interactions generated at order $\alpha^2$, but these become appreciable only where the assumptions of our BCS approach break down, so we ignore them here.
\footnote{Even though order $\alpha^4$ contributions to $\lambda$ (Eq. \ref{lambda}) are small compared to $\lambda$,  they can produce large subleading corrections to  $T_c$ in certain regimes.}
On the other hand, they can give rise to observable effects, 
notably corrections to Fermi liquid theory such as 
quasiparticle mass renormalization, which, while small in the perturbative regime, diverges as a power law approaching the NQCP.

We have also neglected the effects of higher order terms in the effective action, generated at order $\alpha^4$ and beyond. Among others things, these terms include the back-action of the fermions on the  quantum critical dynamics of the nematic modes, {\it i.e.} Landau damping.   
These effect are 
 unimportant so long as $1 \gg \alpha^2 \chi_0 \rho(E_F)$ where $\rho(E_F)$ is  the density of states at the Fermi energy, {\it i.e.} for $|\delta x| \gg \alpha^{2/\gamma}$.  When this inequality is violated, the apparent critical exponents and critical amplitudes that characterize the nematic fluctuations may deviate from their $d+1$ dimensional Ising values at the decoupled NQCP.   

As in the electron-phonon problem, we adopt a perturbative RG approach to account\cite{polchinski} for the retarded nature of $V^{(ind)}$.
We  define  
 dimensionless vertex operators in terms of the 
  interactions and the Fermi velocities, $v_{\hat k}$ (where $\hat k$ denotes a point on the Fermi surface). We then integrate out the Fermionic modes with frequencies between $W$ and $\Omega$.  This results in a new effective action with a high energy cutoff set by $\Omega$, and a renormalized but now instantaneous vertex in the Cooper channel,
\be
{\underline \Gamma} = {\underline \Gamma}^* + {\underline \Gamma}^{(ind)}
\ee
where the 
the non-critical (instantaneous) piece of the vertex operator has been replaced by 
\be
{\underline \Gamma}^* = {\underline \Gamma}^{(0)}[{\underline 1}+ {\underline \Gamma}^{(0)}\log(W/\Omega)]^{-1}
\ee
where $\Gamma^{(0)}_{\hat k,\hat k^\prime}\equiv  V(\vec k,\vec k^\prime)/\sqrt{v_{\hat k}v_{\hat k^\prime}}$.  However, the induced
 interaction is unaffected by this process, 
so
\be
\Gamma^{(ind)}_{\vec k,\vec k^\prime}\approx -\frac{\alpha^2}{4}  \left |f\left(\frac{\vec k+\vec k^{\prime}}{2}\right)\right |^2 \frac{\chi(\vec k-\vec k^\prime,0)}{\sqrt{v_{\hat k}v_{\hat k^\prime}}}.
\ee
This reflects the familiar feature of BCS/Eliashberg theory that only  the instantaneous interaction gets renormalized by the high energy fermionic modes. 

In addition to being  highly peaked at small momentum transfer, {\it i.e.} small $|\vec k-\vec k^\prime|$,  $\Gamma^{(ind)}$ has a significant dependence on the position of $\hat k$ and $\hat k^\prime$ on the Fermi surface: $f(\hat k)$ vanishes at symmetry related ``cold-spots''\cite{Metzner2003} on the Fermi surface, $|k_x|=|k_y|$, and takes on its maximal value, $f(\hat k_{opt})|=1$, at a set of ``optimal pairing points,'' $\hat k_{opt}$. For example for a cuprate-like Fermi surface, these points 
correspond to the ``antinodal 
points'' on the Fermi surface, as illustrated in Fig. 1a. 
Not surprisingly 
 we will find that the strongest pairing occurs for $\vec k$ near $\hat k_{opt}$.

{\bf Solution of the gap equation:}  
We are now left with the problem of fermions with energies within $\Omega$ of the Fermi surface, interacting by an instantaneous interaction vertex $\Gamma$ - i.e. the BCS problem with a $\vec k$ dependent interaction.  Thus, as usual, the superconducting $T_c$ (so long as the weak-coupling condition $T_c\ll \Omega$ is satisfied) is determined as
\be
T_c \sim \Omega \exp[-1/\lambda] 
\label{lambda}
\ee
where in terms of the eigenstates of $\Gamma$,
\be
\int d{\hat k}^\prime \Gamma_{\hat k,\hat k^\prime} \phi^{(a)}_{\hat k^\prime} = -\lambda_{a} \phi^{(a)}_{\hat k}\ ,
\label{eigenstates}
\ee
and $\lambda$ is the largest positive value of $\lambda_a$. 

As a function of $x$, $\Gamma^{(0)}$ is smooth and analytic (neglecting small corrections in the ordered state which we shall discuss), but $\Gamma^{(ind)}_{\hat k,\hat k}$ 
 grows in magnitude upon approach to criticality in proportion to $\alpha^2\chi_0 \rho(E_F) \sim \alpha^2|\delta x|^{-\gamma}$. 
However, 
as we shall see,  the pair wave-function $\phi$ in Eq. \ref {eigenstates} is always a more slowly varying function of $\hat k^\prime$ than is $\Gamma^{(ind)}_{\hat k,\hat k^\prime}$, so the contribution of the induced interactions to $\lambda$ always involves  the integrated weight, 
 \be
 \lambda^{(ind)}
 \equiv \frac{\alpha^2}{4}\int d\hat k v_{\hat k}^{-1}\chi(\hat k,0) \sim \alpha^2   \rho(E_F)\chi_0 (k_F\xi)^{1-d} 
 \label{lambdaind}
 \ee
 where $k_F$ is the Fermi wave-vector.  Therefore, in  $d=2$, $\lambda^{(ind)}$ grows 
 in proportion to $\alpha^2|\delta x|^{\nu-\gamma}$, so the weak coupling BCS approach is valid only for $ |\delta x| > {\cal O}(\alpha^{2/(\gamma - \nu)})$.  However,  in $d=3$, $\gamma-2\nu=0$ so  $\lambda^{(ind)}$ 
 grows only logarithmically, $\lambda^{(ind)}\sim - \alpha^2\log|\delta x|$.   
Our principal remaining task is to analyze the eigenvalue problem in Eq. \ref{eigenstates}.  This is readily done numerically given an explicit form of ${\underline \Gamma}$. 
 We begin, however, by discussing certain limiting cases which can be approximately analyzed analytically.  
 
{\bf Regime of ``weak enhancement'':} The most straightforward regime to analyze
is that  in which the coupling to the nematic mode makes a subdominant contribution to the pairing interaction, 
{\it i.e.} where $\Gamma^{(ind)}$ 
is small compared to $\Gamma^*$. Such a regime always exists sufficiently far from criticality provided that $\lambda^{(0)}\gg \alpha^2$, (where $\lambda^{(0)}$ is the largest positive eigenvalue of $-\Gamma^{(0)}$), a condition we henceforth assume. 

In this regime, the form of the gap function is  largely determined by the non-critical interactions, but $T_c$ is enhanced (possibly by a large factor) by coupling to the nematic modes.  
 This enhancement can be estimated using first order perturbation theory,
\bea
\lambda_a =&& \lambda_a^* + \delta \lambda^{(ind)}_a;  \\ 
\lambda_a^* =&& \lambda_a^{(0)} \ \big\{1-\lambda^{(0)}_a\log[W/\Omega]\big\}^{-1} \nonumber \\
\delta  \lambda^{(ind)}_a=&&-
\int d\hat k  d{\hat k}^\prime \left(\phi^{(a,0)}_{\hat k}\right)^*  \Gamma^{(ind)}_{\hat k,\hat k^\prime} \phi^{(a,0)}_{\hat k^\prime}
\nonumber
\eea
where $\phi^{(a,0)}$ and $\lambda_a^{(0)}$ are, respectively, a (normalized) eigenstate and eigenvalue of ${\underline\Gamma}^{(0)}$.
 In the neighborhood of the NQCP, 
 $\Gamma^{(ind)}$  is peaked  about small $|\hat k-\hat k^\prime|$, hence
\be
 \delta \lambda^{(ind)}_a \approx \lambda^{(ind)} \int d\hat k  |f(\vec k)|^2 \left|\phi^{(a,0)}_{\hat k}\right|^2.
\ee
The degree of the enhancement of pairing thus is larger the more the gap function is peaked 
near $\vec k_{opt}$.
  This result is valid so long as $1 \gg \lambda^{(0)} \gg  \lambda^{(ind)}$. 
  Even so,  the enhancement of $T_c \sim T_c^{(0)}\exp[  \delta\lambda^{(ind)}/(\lambda^{(0)})^2]$ can be large if $  \delta\lambda^{(ind)} \gg [\lambda^{(0)}]^2$, and grows larger the closer one approaches to the NQCP. 
  
 We can also estimate the changes to the form of the gap function, $\Delta_{\vec k}$,  
perturbatively in powers of $\Gamma^{(ind)}$.
 The gap function is  proportional to the pair wave function, $\Delta_{\hat k} \propto \phi_{\hat k}\sqrt{v_{\hat k}}$.  The leading correction to the pair wavefunction is given by 
\be
\phi_{\hat k}\approx\phi_{\hat k}^{(0)}\left[ 1 +\left(\frac { \delta\lambda^{(ind)}} {\lambda^{(0)}}\right) \frac { |f(\vec k)|^2 - \overline { |f|^2}}{\overline {|f|^2}}\right]
\label{wavefunction1}
\ee
where $ \overline { |f|^2}$ is the suitably weighted average of $|f(\vec k)|^2$ over the Fermi surface.
\footnote{The average is  weighted by $|\phi^{(0)}_{\hat k}|^2$, which is by assumption 
a slowly varying function of $\hat k$. }
As a result, the form of the gap function
is little affected by the nematic fluctuations
near the cold spots where $f(\vec k)$ vanishes, but is enhanced far from them.  For example, if $\Delta_{\hat k}^{(0)}$ has the simplest d-wave form, 
$\Delta_{\hat k}^{(0)}\propto [\cos(k_x)-\cos(k_y)]$
, the leading effect of the nematic fluctuations from Eq. \ref{wavefunction1} is to admix an increasing component proportional to 
$(\delta \lambda^{(ind)}/\lambda)[\cos(k_x)-\cos(k_y)]^3$, as seen in Fig. 1b.  In addition, as derived in the supplementary material, the gap 
 is renormalized by $\exp[ \delta \lambda^{(ind)}/(\lambda^{(0)})^2]$ (i.e. the same enhancement factor as $T_c$) compared to its $\alpha=0$ value, but retains a BCS-like $T$ dependence.

{\bf Regime of ``strong enhancement'':}
In 2D, since $\lambda^{(ind)}$ grows rapidly with decreasing $|\delta x|$, there is a crossover to a regime
\footnote{However, for sufficiently large $\lambda^{(0)}$, this crossover occurs closer to criticality than the scale $|\delta x|\sim \alpha^ {2/\gamma}$ at which the boson becomes strongly renormalized}
 in which $\lambda^{(ind)}\gg \lambda^*$. 
In 3D, such a regime
is not generically encountered where our approximations are controlled, 
so we specialize to 2D for 
the present
discussion.
So long as $\lambda^{(ind)} \ll 1$, weak coupling BCS theory still applies, but now the pair wave-function is dominantly determined by $\Gamma^{(ind)}$, while the effects of $\Gamma^*$ can, in turn, be computed perturbatively.  With the cuprates in mind, as illustrated in Fig. 1a, we 
consider a single large closed Fermi surface with four cold spots along the zone diagonals, $|k_x|=|k_y|$, and four optimal points at $\hat k_{opt} = q\hat x + \pi \hat y$ and symmetry related points, although the discussion is readily generalized to more complex Fermi surfaces.  
 
The asymptotic properties of the eigenvalues and eigenstates of $\Gamma^{(ind)}$ can be derived analytically, 
as shown explicitly in the Supplemental Material. The leading eigenstates 
 are peaked about the positions $\hat k_{opt}$
 with an extent in momentum space $\tilde \kappa \sim k_F (\kappa/k_F)^w$, 
 where $w=(1-\eta)/(3-\eta)\approx 1/3$.  Since $w<1$, the eigenstates of $\Gamma^{(ind)}$ vary on a parametrically larger momentum scale than $\Gamma^{(ind )}$ itself, as previously stated.

Since both $\kappa/k_F$ and $\tilde\kappa/k_F \ll 1$, to a first approximation the relative phase of $\phi_{\hat k}$ in the neighborhood of the four optimal points is unimportant, and the eigenfunctions are four-fold degenerate. This degeneracy is lifted by
the large momentum transfer portions of $\Gamma$: 
 The contribution from $\Gamma^{(ind)}$ 
is proportional to $\alpha^2 (\tilde \kappa/k_F)$, which is 
parametrically smaller than $\delta \lambda^*\sim \lambda^* (\tilde \kappa/k_F)$, the perturbative eigenvalue shift 
produced by $\Gamma^*$. 
 Accordingly, the four leading eigenvalues are $\lambda_a \approx \lambda^{(ind)} + g_a \lambda^*(\tilde\kappa/k_F)$ where $|g_a |\sim 1$ depends on the relative phase of the gap function at the different optimal points.  
 Even where the non-critical interactions make a small contribution to the pairing energy, they still determine 
  the relative phase of the pair wave-function at the different optimal points, and hence the symmetry of the superconducting state.
    
For a given symmetry, the splitting between the largest and next-to-largest eigenvalue is of order $\lambda^{(ind)}(\tilde \kappa/k_F)^2\ll \lambda^{(ind)}$. Within the strong enhancement regime, there are several sub-regimes depending on the size of this splitting relative to $[\lambda^{(ind)}]^2$ and 
 to $\delta \lambda^*$.  We defer discussion of 
 sub-regimes to a later paper, but note two salient limits, both within the strong enhancement regime: 1) Sufficiently far from criticality, the form of the gap function
 at all temperatures below $T_c$ is 
 determined by the 
 solution of Eq. \ref{eigenstates};
 even the eccentric shape of the gap function (shown in Figure 1b) results in only modest enhancement of 
 $|\Delta_{max}(T=0)|
 /T_c$.  2) Sufficiently close to criticality, the form of the gap function becomes strongly temperature dependent.  
 In particular, the gap function becomes less strongly peaked at $\hat k_{opt}$ ($\tilde \kappa$ increases) with decreasing $T$. In addition, beyond mean field theory, the near-degeneracy among different symmetry channels within the strong enhancement regime leads to a new class of fluctuations involving the relative phase of the order parameter on different portions of the Fermi surface, as previously explored in Ref. \onlinecite{kunyang2000}.

  {\bf Approaching the NQCP from the ordered phase:}  Until now, we have considered the approach to criticality from the disordered side.  Unlike the case of an antiferromagnetic quantum critical point\cite{bergmetlitski}, in which the opening of a gap on the ordered side of the transition results in a strong suppression of superconductivity, in the case of a NQCP the physics is largely similar when approached from the ordered side.  The major difference is in band structure, i.e. the distortion of the Fermi surface by an amount 
  $\delta k_F\equiv k_{F,x}-k_{F,y} \propto \alpha k_F   \langle\phi\rangle\sim\alpha|\delta x|^{\beta}$.
  $\Gamma^{(ind)}$ is qualitatively affected, because under orthorhombic symmetry there are now only two fermi surface positions $\hat k_{opt}$ of optimal pairing rather than four.  The leading eigenstates of $\Gamma^{(ind)}$ consist of a singlet state of 
  extended s-wave (``$s+d$'')
  symmetry and a triplet state of either $p_x$ or $p_y$ symmetry.
  
  The distortion of the Fermi surface also alters the 
  eigenvalues of both $\Gamma^{*}$ and $\Gamma^{(ind)}$
  by corrections in powers of $\delta k_F$, but these corrections are negligible near criticality. 
The major difference between the ordered and disordered sides comes through the critical amplitude ratio
for the quantity $\chi_0 \xi^{1-d}$.  This is a universal number of order one associated with the decoupled NQCP, and gives the ratio of $\lambda^{(ind)}$ on the two sides of the transition. It is greater than one for $d=2$ and equal to one for $d=3$\cite{Privman1991}, implying that, for fixed $|\delta x|$, $T_c$ is greater on the disordered side in $d=2$ and comparable on both sides in $d=3$.

{\bf Relation to previous work:} The importance of the form factor 
of the coupling between the electrons and 
the quantum critical modes has been explored  in the context of  intra-unit-cell orbital current anti ferromagnetism in Ref. \onlinecite{varma2012}.  However, there the collective modes were  assumed to have an essential $\vec k$ independent susceptibility. The effects 
on Fermi liquids of boson-mediated interactions with strong forward scattering have been treated extensively in various related contexts.\cite{johnston2012,yamase2013,Kulic2000,Rech2006, Dellanna2006,Kim2004,Lee2013}

We were also inspired by two  sets of studies which address superconducting instabilities at a NQCP, Refs. \onlinecite{fitzpatrick}, 
 \onlinecite{fitzpatrick1} and \onlinecite{metlitski}.   Both  address the issue of superconducting pairing asymptotically close to criticality, which is 
the regime we have avoided in the present approach. 
In this regime
the different fields are intrinsically strongly coupled to each other. Thus, in order to obtain theoretical control of the problem, both works involve large $N$ extensions of the model.  
Refs. \onlinecite{fitzpatrick} and \onlinecite{fitzpatrick1} 
 introduce an artificially large number  $N_F$ of fermion flavors and a much larger number of boson flavors, $N_B=(N_F)^2$;  
 no 
 pairing tendency is found to leading order in $1/N_B$ for $d \leq 3$. 
Ref. \onlinecite{metlitski}  treats $d=2$ and $N_B=1$, but extends the  model by introducing both a large  $N_F$ and  a non-local interaction characterized by an exponent, $\epsilon$, assumed small.  
(the physically relevant limit is $N_F=2$, $N_B=1$,
 and $\epsilon=1$.) In contrast to the results of Ref. \onlinecite{fitzpatrick1}, they
  conclude that $T_c$ at criticality is 
 proportional to a power 
 of the coupling constant, 
which is what we would find were we to extrapolate our results to where $\lambda^{(ind)}\sim 1$.

As we were completing this work, we received a paper by Maier and Scalapino\cite{maier2014} reporting a more microscopically realistic study of  the enhancement of $T_c$ by nematic fluctuations - the conclusions are complementary and in broad agreement with the present results.


{\bf Relation to experiment:} 
{The present results provide a rationale to associate the anomalous stability\cite{grissonnanche2014,ramshaw2014} of the superconducting dome in 
near-optimally doped YBCO in high magnetic fields with the proximity of a putative NQCP at doped hole concentration $x \approx 0.18$. The simple $d$-wave (nearly $\cos k_x-\cos k_y$) character of the pairing around this doping, at least in the related material Bi-2212\cite{damascelli-2003}, then suggests that nematic fluctuations play a subdominant role, enhancing a broader tendency to $d$-wave pairing (presumably associated with non-critical magnetic fluctuations).}
The fact that recent evidence indicates that a NQCP occurs at near-optimal doping in some Fe-based superconductors\cite{matsuda2013,analytis2014,fernandes2014} is further evidence that such enhancement may be a more general feature of high temperature superconductivity.  Moreover, the much stronger enhancement of $T_c$ that arises near a NQCP  in 2D may provide some insight as to why $T_c$ is considerably enhanced in  single layer films of 
FeSe.\cite{Wang2012,He2013,Lee2013}

{\bf Acknowledgments:} We thank T. Devereaux, R. Fernandez, I.R. Fisher, E. Fradkin, S. Raghu, B. Ramshaw, and D.J. Scalapino for helpful discussions. This work was supported in part by NSF DMR 1265593 (SAK) and an ABB Fellowship (SL) at Stanford, and by the Israel Science Foundation (\#1291/12) and the Israel-US Binational Science Foundation (\#2012079) at Weizmann (YS and EB).\cite{placeholder}\cite{allendynes,vishik,heLBCO,newvishik,keimer2013}\cite{hill2013,huecker2014,yazdani2014,comin2014}\cite{vishiktrisected,campuzano,Inosov2010,ning2010,stewart,Dong2013,Hirschfeld2011}


\bibliographystyle{apsrev}
\bibliography{Resubmission}

\begin{thebibliography}{59}
\expandafter\ifx\csname natexlab\endcsname\relax\def\natexlab#1{#1}\fi
\expandafter\ifx\csname bibnamefont\endcsname\relax
  \def\bibnamefont#1{#1}\fi
\expandafter\ifx\csname bibfnamefont\endcsname\relax
  \def\bibfnamefont#1{#1}\fi
\expandafter\ifx\csname citenamefont\endcsname\relax
  \def\citenamefont#1{#1}\fi
\expandafter\ifx\csname url\endcsname\relax
  \def\url#1{\texttt{#1}}\fi
\expandafter\ifx\csname urlprefix\endcsname\relax\def\urlprefix{URL }\fi
\providecommand{\bibinfo}[2]{#2}
\providecommand{\eprint}[2][]{\url{#2}}

\bibitem[{\citenamefont{Howald et~al.}(2003)\citenamefont{Howald, Eisaki,
  Kaneko, Greven, and Kapitulnik}}]{howald-2003}
\bibinfo{author}{\bibfnamefont{C.}~\bibnamefont{Howald}},
  \bibinfo{author}{\bibfnamefont{H.}~\bibnamefont{Eisaki}},
  \bibinfo{author}{\bibfnamefont{N.}~\bibnamefont{Kaneko}},
  \bibinfo{author}{\bibfnamefont{M.}~\bibnamefont{Greven}}, \bibnamefont{and}
  \bibinfo{author}{\bibfnamefont{A.}~\bibnamefont{Kapitulnik}},
  \bibinfo{journal}{Phys. Rev. B} \textbf{\bibinfo{volume}{67}},
  \bibinfo{pages}{014533} (\bibinfo{year}{2003}).

\bibitem[{\citenamefont{Xia et~al.}(2008)\citenamefont{Xia, Schemm, Deutscher,
  Kivelson, Bonn, Hardy, Liang, Siemons, Koster, Fejer et~al.}}]{xia-2007}
\bibinfo{author}{\bibfnamefont{J.}~\bibnamefont{Xia}},
  \bibinfo{author}{\bibfnamefont{E.}~\bibnamefont{Schemm}},
  \bibinfo{author}{\bibfnamefont{G.}~\bibnamefont{Deutscher}},
  \bibinfo{author}{\bibfnamefont{S.~A.} \bibnamefont{Kivelson}},
  \bibinfo{author}{\bibfnamefont{D.~A.} \bibnamefont{Bonn}},
  \bibinfo{author}{\bibfnamefont{W.~N.} \bibnamefont{Hardy}},
  \bibinfo{author}{\bibfnamefont{R.}~\bibnamefont{Liang}},
  \bibinfo{author}{\bibfnamefont{W.}~\bibnamefont{Siemons}},
  \bibinfo{author}{\bibfnamefont{G.}~\bibnamefont{Koster}},
  \bibinfo{author}{\bibfnamefont{M.~M.} \bibnamefont{Fejer}},
  \bibnamefont{et~al.}, \bibinfo{journal}{Phys. Rev. Lett.}
  \textbf{\bibinfo{volume}{100}}, \bibinfo{pages}{127002}
  (\bibinfo{year}{2008}).

\bibitem[{\citenamefont{Hinkov et~al.}(2008)\citenamefont{Hinkov, Haug,
  Fauqu{\'e}, Bourges, Sidis, Ivanov, Bernhard, Lin, and Keimer}}]{hinkov-2007}
\bibinfo{author}{\bibfnamefont{V.}~\bibnamefont{Hinkov}},
  \bibinfo{author}{\bibfnamefont{D.}~\bibnamefont{Haug}},
  \bibinfo{author}{\bibfnamefont{B.}~\bibnamefont{Fauqu{\'e}}},
  \bibinfo{author}{\bibfnamefont{P.}~\bibnamefont{Bourges}},
  \bibinfo{author}{\bibfnamefont{Y.}~\bibnamefont{Sidis}},
  \bibinfo{author}{\bibfnamefont{A.}~\bibnamefont{Ivanov}},
  \bibinfo{author}{\bibfnamefont{C.}~\bibnamefont{Bernhard}},
  \bibinfo{author}{\bibfnamefont{C.~T.} \bibnamefont{Lin}}, \bibnamefont{and}
  \bibinfo{author}{\bibfnamefont{B.}~\bibnamefont{Keimer}},
  \bibinfo{journal}{Science} \textbf{\bibinfo{volume}{319}},
  \bibinfo{pages}{597} (\bibinfo{year}{2008}).

\bibitem[{\citenamefont{Daou et~al.}(2010)\citenamefont{Daou, Chang, LeBoeuf,
  Cyr-Choini{\`e}re, Lalibert{\'e}, Doiron-Leyraud, Ramshaw, Liang, Bonn, Hardy
  et~al.}}]{taillefer-nematic-2009}
\bibinfo{author}{\bibfnamefont{R.}~\bibnamefont{Daou}},
  \bibinfo{author}{\bibfnamefont{J.}~\bibnamefont{Chang}},
  \bibinfo{author}{\bibfnamefont{D.}~\bibnamefont{LeBoeuf}},
  \bibinfo{author}{\bibfnamefont{O.}~\bibnamefont{Cyr-Choini{\`e}re}},
  \bibinfo{author}{\bibfnamefont{F.}~\bibnamefont{Lalibert{\'e}}},
  \bibinfo{author}{\bibfnamefont{N.}~\bibnamefont{Doiron-Leyraud}},
  \bibinfo{author}{\bibfnamefont{B.~J.} \bibnamefont{Ramshaw}},
  \bibinfo{author}{\bibfnamefont{R.}~\bibnamefont{Liang}},
  \bibinfo{author}{\bibfnamefont{D.~A.} \bibnamefont{Bonn}},
  \bibinfo{author}{\bibfnamefont{W.~N.} \bibnamefont{Hardy}},
  \bibnamefont{et~al.}, \bibinfo{journal}{Nature}
  \textbf{\bibinfo{volume}{463}}, \bibinfo{pages}{519} (\bibinfo{year}{2010}).

\bibitem[{\citenamefont{Lawler et~al.}(2010)\citenamefont{Lawler, Fujita, Lee,
  Schmidt, Kohsaka, Kim, Eisaki, Uchida, Davis, Sethna et~al.}}]{lawler-2010}
\bibinfo{author}{\bibfnamefont{M.~J.} \bibnamefont{Lawler}},
  \bibinfo{author}{\bibfnamefont{K.}~\bibnamefont{Fujita}},
  \bibinfo{author}{\bibfnamefont{J.~W.} \bibnamefont{Lee}},
  \bibinfo{author}{\bibfnamefont{A.~R.} \bibnamefont{Schmidt}},
  \bibinfo{author}{\bibfnamefont{Y.}~\bibnamefont{Kohsaka}},
  \bibinfo{author}{\bibfnamefont{C.~K.} \bibnamefont{Kim}},
  \bibinfo{author}{\bibfnamefont{H.}~\bibnamefont{Eisaki}},
  \bibinfo{author}{\bibfnamefont{S.}~\bibnamefont{Uchida}},
  \bibinfo{author}{\bibfnamefont{J.~C.} \bibnamefont{Davis}},
  \bibinfo{author}{\bibfnamefont{J.~P.} \bibnamefont{Sethna}},
  \bibnamefont{et~al.}, \bibinfo{journal}{Nature}
  \textbf{\bibinfo{volume}{466}}, \bibinfo{pages}{347} (\bibinfo{year}{2010}).

\bibitem[{\citenamefont{Wu et~al.}(2014)\citenamefont{Wu, Mayaffre, Kr{\"a}mer,
  Horvati{\'c}, Berthier, Hardy, Liang, Bonn, and Julien}}]{wu-2014}
\bibinfo{author}{\bibfnamefont{T.}~\bibnamefont{Wu}},
  \bibinfo{author}{\bibfnamefont{H.}~\bibnamefont{Mayaffre}},
  \bibinfo{author}{\bibfnamefont{S.}~\bibnamefont{Kr{\"a}mer}},
  \bibinfo{author}{\bibfnamefont{M.}~\bibnamefont{Horvati{\'c}}},
  \bibinfo{author}{\bibfnamefont{C.}~\bibnamefont{Berthier}},
  \bibinfo{author}{\bibfnamefont{W.}~\bibnamefont{Hardy}},
  \bibinfo{author}{\bibfnamefont{R.}~\bibnamefont{Liang}},
  \bibinfo{author}{\bibfnamefont{D.}~\bibnamefont{Bonn}}, \bibnamefont{and}
  \bibinfo{author}{\bibfnamefont{M.-H.} \bibnamefont{Julien}}
  (\bibinfo{year}{2014}), \bibinfo{note}{unpublished},
  \eprint{arXiv:1404.1617}.

\bibitem[{\citenamefont{Grissonnanche et~al.}(2014)\citenamefont{Grissonnanche,
  Cyr-Choinire, LalibertŽ, RenŽde~Cotret, Juneau-Fecteau, Dufour-BeausŽjour,
  Delage, LeBoeuf, Chang, Ramshaw et~al.}}]{grissonnanche2014}
\bibinfo{author}{\bibfnamefont{G.}~\bibnamefont{Grissonnanche}},
  \bibinfo{author}{\bibfnamefont{O.}~\bibnamefont{Cyr-Choinire}},
  \bibinfo{author}{\bibfnamefont{F.}~\bibnamefont{LalibertŽ}},
  \bibinfo{author}{\bibfnamefont{S.}~\bibnamefont{RenŽde~Cotret}},
  \bibinfo{author}{\bibfnamefont{A.}~\bibnamefont{Juneau-Fecteau}},
  \bibinfo{author}{\bibfnamefont{S.}~\bibnamefont{Dufour-BeausŽjour}},
  \bibinfo{author}{\bibfnamefont{M.~é.} \bibnamefont{Delage}},
  \bibinfo{author}{\bibfnamefont{D.}~\bibnamefont{LeBoeuf}},
  \bibinfo{author}{\bibfnamefont{J.}~\bibnamefont{Chang}},
  \bibinfo{author}{\bibfnamefont{B.~J.} \bibnamefont{Ramshaw}},
  \bibnamefont{et~al.}, \bibinfo{journal}{Nat Commun}
  \textbf{\bibinfo{volume}{5}}, \bibinfo{pages}{3280} (\bibinfo{year}{2014}).

\bibitem[{\citenamefont{Ramshaw et~al.}(2014)\citenamefont{Ramshaw, Sebastian,
  McDonald, Day, Tam, Zhu, Betts, Liang, Bonn, Hardy et~al.}}]{ramshaw2014}
\bibinfo{author}{\bibfnamefont{B.~J.} \bibnamefont{Ramshaw}},
  \bibinfo{author}{\bibfnamefont{S.}~\bibnamefont{Sebastian}},
  \bibinfo{author}{\bibfnamefont{R.}~\bibnamefont{McDonald}},
  \bibinfo{author}{\bibfnamefont{J.}~\bibnamefont{Day}},
  \bibinfo{author}{\bibfnamefont{B.}~\bibnamefont{Tam}},
  \bibinfo{author}{\bibfnamefont{Z.}~\bibnamefont{Zhu}},
  \bibinfo{author}{\bibfnamefont{J.}~\bibnamefont{Betts}},
  \bibinfo{author}{\bibfnamefont{R.}~\bibnamefont{Liang}},
  \bibinfo{author}{\bibfnamefont{D.}~\bibnamefont{Bonn}},
  \bibinfo{author}{\bibfnamefont{W.~N.} \bibnamefont{Hardy}},
  \bibnamefont{et~al.} (\bibinfo{year}{2014}), \eprint{1405.5238}.

\bibitem[{\citenamefont{{Fujita} et~al.}(2014)\citenamefont{{Fujita}, {Kim},
  {Lee}, {Lee}, {Hamidian}, {Firmo}, {Mukhopadhyay}, {Eisaki}, {Uchida},
  {Lawler} et~al.}}]{fujita2014}
\bibinfo{author}{\bibfnamefont{K.}~\bibnamefont{{Fujita}}},
  \bibinfo{author}{\bibfnamefont{C.~K.} \bibnamefont{{Kim}}},
  \bibinfo{author}{\bibfnamefont{I.}~\bibnamefont{{Lee}}},
  \bibinfo{author}{\bibfnamefont{J.}~\bibnamefont{{Lee}}},
  \bibinfo{author}{\bibfnamefont{M.~H.} \bibnamefont{{Hamidian}}},
  \bibinfo{author}{\bibfnamefont{I.~A.} \bibnamefont{{Firmo}}},
  \bibinfo{author}{\bibfnamefont{S.}~\bibnamefont{{Mukhopadhyay}}},
  \bibinfo{author}{\bibfnamefont{H.}~\bibnamefont{{Eisaki}}},
  \bibinfo{author}{\bibfnamefont{S.}~\bibnamefont{{Uchida}}},
  \bibinfo{author}{\bibfnamefont{M.~J.} \bibnamefont{{Lawler}}},
  \bibnamefont{et~al.}, \bibinfo{journal}{ArXiv e-prints}
  (\bibinfo{year}{2014}), \eprint{1403.7788}.

\bibitem[{\citenamefont{Chu et~al.}(2010)\citenamefont{Chu, Analytis, {De
  Greve}, McMahon, Islam, Yamamoto, and Fisher}}]{fisher-2010}
\bibinfo{author}{\bibfnamefont{J.-H.} \bibnamefont{Chu}},
  \bibinfo{author}{\bibfnamefont{J.~G.} \bibnamefont{Analytis}},
  \bibinfo{author}{\bibfnamefont{K.}~\bibnamefont{{De Greve}}},
  \bibinfo{author}{\bibfnamefont{P.~L.} \bibnamefont{McMahon}},
  \bibinfo{author}{\bibfnamefont{Z.}~\bibnamefont{Islam}},
  \bibinfo{author}{\bibfnamefont{Y.}~\bibnamefont{Yamamoto}}, \bibnamefont{and}
  \bibinfo{author}{\bibfnamefont{I.~R.} \bibnamefont{Fisher}},
  \bibinfo{journal}{Science} \textbf{\bibinfo{volume}{329}},
  \bibinfo{pages}{824} (\bibinfo{year}{2010}).

\bibitem[{\citenamefont{{Chu} et~al.}(2012)\citenamefont{{Chu}, {Kuo},
  {Analytis}, and {Fisher}}}]{chu2012}
\bibinfo{author}{\bibfnamefont{J.-H.} \bibnamefont{{Chu}}},
  \bibinfo{author}{\bibfnamefont{H.-H.} \bibnamefont{{Kuo}}},
  \bibinfo{author}{\bibfnamefont{J.~G.} \bibnamefont{{Analytis}}},
  \bibnamefont{and} \bibinfo{author}{\bibfnamefont{I.~R.}
  \bibnamefont{{Fisher}}}, \bibinfo{journal}{Science}
  \textbf{\bibinfo{volume}{337}}, \bibinfo{pages}{710} (\bibinfo{year}{2012}),
  \eprint{1203.3239}.

\bibitem[{\citenamefont{{Yoshizawa} et~al.}(2012)\citenamefont{{Yoshizawa},
  {Kimura}, {Chiba}, {Simayi}, {Nakanishi}, {Kihou}, {Lee}, {Iyo}, {Eisaki},
  {Nakajima} et~al.}}]{yoshizawa2012}
\bibinfo{author}{\bibfnamefont{M.}~\bibnamefont{{Yoshizawa}}},
  \bibinfo{author}{\bibfnamefont{D.}~\bibnamefont{{Kimura}}},
  \bibinfo{author}{\bibfnamefont{T.}~\bibnamefont{{Chiba}}},
  \bibinfo{author}{\bibfnamefont{S.}~\bibnamefont{{Simayi}}},
  \bibinfo{author}{\bibfnamefont{Y.}~\bibnamefont{{Nakanishi}}},
  \bibinfo{author}{\bibfnamefont{K.}~\bibnamefont{{Kihou}}},
  \bibinfo{author}{\bibfnamefont{C.-H.} \bibnamefont{{Lee}}},
  \bibinfo{author}{\bibfnamefont{A.}~\bibnamefont{{Iyo}}},
  \bibinfo{author}{\bibfnamefont{H.}~\bibnamefont{{Eisaki}}},
  \bibinfo{author}{\bibfnamefont{M.}~\bibnamefont{{Nakajima}}},
  \bibnamefont{et~al.}, \bibinfo{journal}{Journal of the Physical Society of
  Japan} \textbf{\bibinfo{volume}{81}}, \bibinfo{pages}{024604}
  (\bibinfo{year}{2012}), \eprint{1111.0366}.

\bibitem[{\citenamefont{Walmsley et~al.}({2013})\citenamefont{Walmsley, Putzke,
  Malone, Guillamon, Vignolles, Proust, Badoux, Coldea, Watson, Kasahara
  et~al.}}]{matsuda2013}
\bibinfo{author}{\bibfnamefont{P.}~\bibnamefont{Walmsley}},
  \bibinfo{author}{\bibfnamefont{C.}~\bibnamefont{Putzke}},
  \bibinfo{author}{\bibfnamefont{L.}~\bibnamefont{Malone}},
  \bibinfo{author}{\bibfnamefont{I.}~\bibnamefont{Guillamon}},
  \bibinfo{author}{\bibfnamefont{D.}~\bibnamefont{Vignolles}},
  \bibinfo{author}{\bibfnamefont{C.}~\bibnamefont{Proust}},
  \bibinfo{author}{\bibfnamefont{S.}~\bibnamefont{Badoux}},
  \bibinfo{author}{\bibfnamefont{A.~I.} \bibnamefont{Coldea}},
  \bibinfo{author}{\bibfnamefont{M.~D.} \bibnamefont{Watson}},
  \bibinfo{author}{\bibfnamefont{S.}~\bibnamefont{Kasahara}},
  \bibnamefont{et~al.}, \bibinfo{journal}{{PHYSICAL REVIEW LETTERS}}
  \textbf{\bibinfo{volume}{{110}}}, \bibinfo{pages}{257002}
  (\bibinfo{year}{{2013}}).

\bibitem[{\citenamefont{Berg et~al.}(2009)\citenamefont{Berg, Fradkin,
  Kivelson, and Tranquada}}]{berg-2009b}
\bibinfo{author}{\bibfnamefont{E.}~\bibnamefont{Berg}},
  \bibinfo{author}{\bibfnamefont{E.}~\bibnamefont{Fradkin}},
  \bibinfo{author}{\bibfnamefont{S.~A.} \bibnamefont{Kivelson}},
  \bibnamefont{and} \bibinfo{author}{\bibfnamefont{J.~M.}
  \bibnamefont{Tranquada}}, \bibinfo{journal}{New J. Phys.}
  \textbf{\bibinfo{volume}{11}}, \bibinfo{pages}{115004}
  (\bibinfo{year}{2009}).

\bibitem[{\citenamefont{Fradkin and Kivelson}(2012)}]{Fradkin-2012}
\bibinfo{author}{\bibfnamefont{E.}~\bibnamefont{Fradkin}} \bibnamefont{and}
  \bibinfo{author}{\bibfnamefont{S.~A.} \bibnamefont{Kivelson}},
  \bibinfo{journal}{Nat. Phys.} \textbf{\bibinfo{volume}{8}},
  \bibinfo{pages}{864} (\bibinfo{year}{2012}).

\bibitem[{\citenamefont{{S{\'e}amus Davis} and {Lee}}(2013)}]{davis2013}
\bibinfo{author}{\bibfnamefont{J.~C.} \bibnamefont{{S{\'e}amus Davis}}}
  \bibnamefont{and} \bibinfo{author}{\bibfnamefont{D.-H.} \bibnamefont{{Lee}}},
  \bibinfo{journal}{ArXiv e-prints}  (\bibinfo{year}{2013}),
  \eprint{1309.2719}.

\bibitem[{\citenamefont{{Lee}}(2014)}]{palee2014}
\bibinfo{author}{\bibfnamefont{P.~A.} \bibnamefont{{Lee}}},
  \bibinfo{journal}{ArXiv e-prints}  (\bibinfo{year}{2014}),
  \eprint{1401.0519}.

\bibitem[{\citenamefont{Nie et~al.}(2014)\citenamefont{Nie, Tarjus, and
  Kivelson}}]{nie-2013}
\bibinfo{author}{\bibfnamefont{L.}~\bibnamefont{Nie}},
  \bibinfo{author}{\bibfnamefont{G.}~\bibnamefont{Tarjus}}, \bibnamefont{and}
  \bibinfo{author}{\bibfnamefont{S.~A.} \bibnamefont{Kivelson}},
  \bibinfo{journal}{Proceedings of the National Academy of Sciences}
  \textbf{\bibinfo{volume}{111}}, \bibinfo{pages}{7980} (\bibinfo{year}{2014}).

\bibitem[{\citenamefont{Shankar}(1994)}]{shankar}
\bibinfo{author}{\bibfnamefont{R.}~\bibnamefont{Shankar}},
  \bibinfo{journal}{Rev. Mod. Phys.} \textbf{\bibinfo{volume}{66}},
  \bibinfo{pages}{129} (\bibinfo{year}{1994}).

\bibitem[{\citenamefont{Polchinski}(1992)}]{polchinski}
\bibinfo{author}{\bibfnamefont{J.}~\bibnamefont{Polchinski}}
  (\bibinfo{year}{1992}), \eprint{9210046v2}.

\bibitem[{\citenamefont{Dahm et~al.}(2009)\citenamefont{Dahm, Hinkov,
  Borisenko, Kordyuk, Zabolotnyy, Fink, B\"{u}chner, Scalapino, Hanke, and
  Keimer}}]{keimer}
\bibinfo{author}{\bibfnamefont{T.}~\bibnamefont{Dahm}},
  \bibinfo{author}{\bibfnamefont{V.}~\bibnamefont{Hinkov}},
  \bibinfo{author}{\bibfnamefont{S.~V.} \bibnamefont{Borisenko}},
  \bibinfo{author}{\bibfnamefont{a.~a.} \bibnamefont{Kordyuk}},
  \bibinfo{author}{\bibfnamefont{V.~B.} \bibnamefont{Zabolotnyy}},
  \bibinfo{author}{\bibfnamefont{J.}~\bibnamefont{Fink}},
  \bibinfo{author}{\bibfnamefont{B.}~\bibnamefont{B\"{u}chner}},
  \bibinfo{author}{\bibfnamefont{D.~J.} \bibnamefont{Scalapino}},
  \bibinfo{author}{\bibfnamefont{W.}~\bibnamefont{Hanke}}, \bibnamefont{and}
  \bibinfo{author}{\bibfnamefont{B.}~\bibnamefont{Keimer}},
  \bibinfo{journal}{Nat. Phys.} \textbf{\bibinfo{volume}{5}},
  \bibinfo{pages}{217} (\bibinfo{year}{2009}).

\bibitem[{\citenamefont{Privman et~al.}(1991)\citenamefont{Privman, Hohenberg,
  and Aharony}}]{Privman1991}
\bibinfo{author}{\bibfnamefont{V.}~\bibnamefont{Privman}},
  \bibinfo{author}{\bibfnamefont{P.}~\bibnamefont{Hohenberg}},
  \bibnamefont{and} \bibinfo{author}{\bibfnamefont{A.}~\bibnamefont{Aharony}},
  in \emph{\bibinfo{booktitle}{Phase Transitions and Critical Phenomena, Vol.
  14}}, edited by \bibinfo{editor}{\bibfnamefont{C.}~\bibnamefont{Domb}}
  \bibnamefont{and} \bibinfo{editor}{\bibfnamefont{J.}~\bibnamefont{Lebowitz}}
  (\bibinfo{publisher}{Academic Press}, \bibinfo{address}{New York},
  \bibinfo{year}{1991}).

\bibitem[{\citenamefont{Metzner et~al.}(2003)\citenamefont{Metzner, Rohe, and
  Andergassen}}]{Metzner2003}
\bibinfo{author}{\bibfnamefont{W.}~\bibnamefont{Metzner}},
  \bibinfo{author}{\bibfnamefont{D.}~\bibnamefont{Rohe}}, \bibnamefont{and}
  \bibinfo{author}{\bibfnamefont{S.}~\bibnamefont{Andergassen}},
  \bibinfo{journal}{Phys. Rev. Lett.} \textbf{\bibinfo{volume}{91}},
  \bibinfo{pages}{066402} (\bibinfo{year}{2003}), ISSN
  \bibinfo{issn}{0031-9007}.

\bibitem[{\citenamefont{Yang and Sondhi}(2000)}]{kunyang2000}
\bibinfo{author}{\bibfnamefont{K.}~\bibnamefont{Yang}} \bibnamefont{and}
  \bibinfo{author}{\bibfnamefont{S.~L.} \bibnamefont{Sondhi}},
  \bibinfo{journal}{Phys. Rev. B} \textbf{\bibinfo{volume}{62}},
  \bibinfo{pages}{11778} (\bibinfo{year}{2000}).

\bibitem[{\citenamefont{Berg et~al.}(2012)\citenamefont{Berg, Metlitski, and
  Sachdev}}]{bergmetlitski}
\bibinfo{author}{\bibfnamefont{E.}~\bibnamefont{Berg}},
  \bibinfo{author}{\bibfnamefont{M.~A.} \bibnamefont{Metlitski}},
  \bibnamefont{and} \bibinfo{author}{\bibfnamefont{S.}~\bibnamefont{Sachdev}},
  \bibinfo{journal}{SCIENCE} \textbf{\bibinfo{volume}{338}},
  \bibinfo{pages}{1606} (\bibinfo{year}{2012}).

\bibitem[{\citenamefont{Varma}({2012})}]{varma2012}
\bibinfo{author}{\bibfnamefont{C.~M.} \bibnamefont{Varma}},
  \bibinfo{journal}{{Reports on Progress in Pysics}}
  \textbf{\bibinfo{volume}{{75}}}, \bibinfo{pages}{052501}
  (\bibinfo{year}{{2012}}).

\bibitem[{\citenamefont{Johnston et~al.}({2012})\citenamefont{Johnston, Vishik,
  Lee, Schmitt, Uchida, Fujita, Ishida, Nagaosa, Shen, and
  Devereaux}}]{johnston2012}
\bibinfo{author}{\bibfnamefont{S.}~\bibnamefont{Johnston}},
  \bibinfo{author}{\bibfnamefont{I.~M.} \bibnamefont{Vishik}},
  \bibinfo{author}{\bibfnamefont{W.~S.} \bibnamefont{Lee}},
  \bibinfo{author}{\bibfnamefont{F.}~\bibnamefont{Schmitt}},
  \bibinfo{author}{\bibfnamefont{S.}~\bibnamefont{Uchida}},
  \bibinfo{author}{\bibfnamefont{K.}~\bibnamefont{Fujita}},
  \bibinfo{author}{\bibfnamefont{S.}~\bibnamefont{Ishida}},
  \bibinfo{author}{\bibfnamefont{N.}~\bibnamefont{Nagaosa}},
  \bibinfo{author}{\bibfnamefont{Z.~X.} \bibnamefont{Shen}}, \bibnamefont{and}
  \bibinfo{author}{\bibfnamefont{T.~P.} \bibnamefont{Devereaux}},
  \bibinfo{journal}{{PHYSICAL REVIEW LETTERS}} \textbf{\bibinfo{volume}{{108}}}
  (\bibinfo{year}{{2012}}).

\bibitem[{\citenamefont{Yamase and Zeyher}(2013)}]{yamase2013}
\bibinfo{author}{\bibfnamefont{H.}~\bibnamefont{Yamase}} \bibnamefont{and}
  \bibinfo{author}{\bibfnamefont{R.}~\bibnamefont{Zeyher}},
  \bibinfo{journal}{Phys. Rev. B} \textbf{\bibinfo{volume}{88}},
  \bibinfo{pages}{180502} (\bibinfo{year}{2013}).

\bibitem[{\citenamefont{Kulic}(2000)}]{Kulic2000}
\bibinfo{author}{\bibfnamefont{M.~L.} \bibnamefont{Kulic}},
  \bibinfo{journal}{Phys. Rep.} \textbf{\bibinfo{volume}{338}},
  \bibinfo{pages}{1} (\bibinfo{year}{2000}).

\bibitem[{\citenamefont{Rech et~al.}(2006)\citenamefont{Rech, P\'{e}pin, and
  Chubukov}}]{Rech2006}
\bibinfo{author}{\bibfnamefont{J.}~\bibnamefont{Rech}},
  \bibinfo{author}{\bibfnamefont{C.}~\bibnamefont{P\'{e}pin}},
  \bibnamefont{and} \bibinfo{author}{\bibfnamefont{A.}~\bibnamefont{Chubukov}},
  \bibinfo{journal}{Phys. Rev. B} \textbf{\bibinfo{volume}{74}},
  \bibinfo{pages}{195126} (\bibinfo{year}{2006}).

\bibitem[{\citenamefont{Dell'Anna and Metzner}(2006)}]{Dellanna2006}
\bibinfo{author}{\bibfnamefont{L.}~\bibnamefont{Dell'Anna}} \bibnamefont{and}
  \bibinfo{author}{\bibfnamefont{W.}~\bibnamefont{Metzner}},
  \bibinfo{journal}{Phys. Rev. B} \textbf{\bibinfo{volume}{73}},
  \bibinfo{pages}{045127} (\bibinfo{year}{2006}).

\bibitem[{\citenamefont{Kim and Kee}(2004)}]{Kim2004}
\bibinfo{author}{\bibfnamefont{Y.~B.} \bibnamefont{Kim}} \bibnamefont{and}
  \bibinfo{author}{\bibfnamefont{H.-Y.} \bibnamefont{Kee}},
  \bibinfo{journal}{Journal of Physics: Condensed Matter}
  \textbf{\bibinfo{volume}{16}}, \bibinfo{pages}{3139} (\bibinfo{year}{2004}).

\bibitem[{\citenamefont{{Lee} et~al.}(2013)\citenamefont{{Lee}, {Schmitt},
  {Moore}, {Johnston}, {Cui}, {Li}, {Yi}, {Liu}, {Hashimoto}, {Zhang}
  et~al.}}]{Lee2013}
\bibinfo{author}{\bibfnamefont{J.~J.} \bibnamefont{{Lee}}},
  \bibinfo{author}{\bibfnamefont{F.~T.} \bibnamefont{{Schmitt}}},
  \bibinfo{author}{\bibfnamefont{R.~G.} \bibnamefont{{Moore}}},
  \bibinfo{author}{\bibfnamefont{S.}~\bibnamefont{{Johnston}}},
  \bibinfo{author}{\bibfnamefont{Y.-T.} \bibnamefont{{Cui}}},
  \bibinfo{author}{\bibfnamefont{W.}~\bibnamefont{{Li}}},
  \bibinfo{author}{\bibfnamefont{M.}~\bibnamefont{{Yi}}},
  \bibinfo{author}{\bibfnamefont{Z.~K.} \bibnamefont{{Liu}}},
  \bibinfo{author}{\bibfnamefont{M.}~\bibnamefont{{Hashimoto}}},
  \bibinfo{author}{\bibfnamefont{Y.}~\bibnamefont{{Zhang}}},
  \bibnamefont{et~al.}, \bibinfo{journal}{ArXiv e-prints}
  (\bibinfo{year}{2013}), \eprint{1312.2633}.

\bibitem[{\citenamefont{{Fitzpatrick} et~al.}(2014)\citenamefont{{Fitzpatrick},
  {Kachru}, {Kaplan}, and {Raghu}}}]{fitzpatrick}
\bibinfo{author}{\bibfnamefont{A.~L.} \bibnamefont{{Fitzpatrick}}},
  \bibinfo{author}{\bibfnamefont{S.}~\bibnamefont{{Kachru}}},
  \bibinfo{author}{\bibfnamefont{J.}~\bibnamefont{{Kaplan}}}, \bibnamefont{and}
  \bibinfo{author}{\bibfnamefont{S.}~\bibnamefont{{Raghu}}},
  \bibinfo{journal}{\prb} \textbf{\bibinfo{volume}{89}}, \bibinfo{eid}{165114}
  (\bibinfo{year}{2014}), \eprint{1312.3321}.

\bibitem[{\citenamefont{Fitzpatrick et~al.}(2013)\citenamefont{Fitzpatrick,
  Kachru, Kaplan, and Raghu}}]{fitzpatrick1}
\bibinfo{author}{\bibfnamefont{A.~L.} \bibnamefont{Fitzpatrick}},
  \bibinfo{author}{\bibfnamefont{S.}~\bibnamefont{Kachru}},
  \bibinfo{author}{\bibfnamefont{J.}~\bibnamefont{Kaplan}}, \bibnamefont{and}
  \bibinfo{author}{\bibfnamefont{S.}~\bibnamefont{Raghu}},
  \bibinfo{journal}{Phys. Rev. B} \textbf{\bibinfo{volume}{88}},
  \bibinfo{pages}{125116} (\bibinfo{year}{2013}).

\bibitem[{\citenamefont{{Metlitski} et~al.}(2014)\citenamefont{{Metlitski},
  {Mross}, {Sachdev}, and {Senthil}}}]{metlitski}
\bibinfo{author}{\bibfnamefont{M.~A.} \bibnamefont{{Metlitski}}},
  \bibinfo{author}{\bibfnamefont{D.~F.} \bibnamefont{{Mross}}},
  \bibinfo{author}{\bibfnamefont{S.}~\bibnamefont{{Sachdev}}},
  \bibnamefont{and}
  \bibinfo{author}{\bibfnamefont{T.}~\bibnamefont{{Senthil}}},
  \bibinfo{journal}{ArXiv e-prints}  (\bibinfo{year}{2014}),
  \eprint{1403.3694}.

\bibitem[{\citenamefont{Maier and Scalapino}(2014)}]{maier2014}
\bibinfo{author}{\bibfnamefont{T.}~\bibnamefont{Maier}} \bibnamefont{and}
  \bibinfo{author}{\bibfnamefont{D.~J.} \bibnamefont{Scalapino}}
  (\bibinfo{year}{2014}), \eprint{1405.5238}.

\bibitem[{\citenamefont{Damascelli et~al.}(2003)\citenamefont{Damascelli, Shen,
  and Hussain}}]{damascelli-2003}
\bibinfo{author}{\bibfnamefont{A.}~\bibnamefont{Damascelli}},
  \bibinfo{author}{\bibfnamefont{Z.-X.} \bibnamefont{Shen}}, \bibnamefont{and}
  \bibinfo{author}{\bibfnamefont{Z.}~\bibnamefont{Hussain}},
  \bibinfo{journal}{Rev. Mod. Phys.} \textbf{\bibinfo{volume}{75}},
  \bibinfo{pages}{473} (\bibinfo{year}{2003}).

\bibitem[{\citenamefont{Analytis et~al.}(2014)\citenamefont{Analytis, Kuo,
  McDonald, Wartenbe, Rourke, Hussey, and Fisher}}]{analytis2014}
\bibinfo{author}{\bibfnamefont{J.~G.} \bibnamefont{Analytis}},
  \bibinfo{author}{\bibfnamefont{H.-H.} \bibnamefont{Kuo}},
  \bibinfo{author}{\bibfnamefont{R.~D.} \bibnamefont{McDonald}},
  \bibinfo{author}{\bibfnamefont{M.}~\bibnamefont{Wartenbe}},
  \bibinfo{author}{\bibfnamefont{P.~M.~C.} \bibnamefont{Rourke}},
  \bibinfo{author}{\bibfnamefont{N.~E.} \bibnamefont{Hussey}},
  \bibnamefont{and} \bibinfo{author}{\bibfnamefont{I.~R.}
  \bibnamefont{Fisher}}, \bibinfo{journal}{NATURE PHYSICS}
  \textbf{\bibinfo{volume}{10}}, \bibinfo{pages}{194} (\bibinfo{year}{2014}).

\bibitem[{\citenamefont{Fernandes et~al.}(2014)\citenamefont{Fernandes,
  Chubukov, and Schmalian}}]{fernandes2014}
\bibinfo{author}{\bibfnamefont{R.~M.} \bibnamefont{Fernandes}},
  \bibinfo{author}{\bibfnamefont{A.~V.} \bibnamefont{Chubukov}},
  \bibnamefont{and}
  \bibinfo{author}{\bibfnamefont{J.}~\bibnamefont{Schmalian}},
  \bibinfo{journal}{NATURE PHYSICS} \textbf{\bibinfo{volume}{10}},
  \bibinfo{pages}{97} (\bibinfo{year}{2014}).

\bibitem[{\citenamefont{Wang et~al.}(2012)\citenamefont{Wang, Li, Zhang, Zhang,
  Zhang, Li, Ding, Ou, Deng, Chang et~al.}}]{Wang2012}
\bibinfo{author}{\bibfnamefont{Q.-Y.} \bibnamefont{Wang}},
  \bibinfo{author}{\bibfnamefont{Z.}~\bibnamefont{Li}},
  \bibinfo{author}{\bibfnamefont{W.-H.} \bibnamefont{Zhang}},
  \bibinfo{author}{\bibfnamefont{Z.-C.} \bibnamefont{Zhang}},
  \bibinfo{author}{\bibfnamefont{J.-S.} \bibnamefont{Zhang}},
  \bibinfo{author}{\bibfnamefont{W.}~\bibnamefont{Li}},
  \bibinfo{author}{\bibfnamefont{H.}~\bibnamefont{Ding}},
  \bibinfo{author}{\bibfnamefont{Y.-B.} \bibnamefont{Ou}},
  \bibinfo{author}{\bibfnamefont{P.}~\bibnamefont{Deng}},
  \bibinfo{author}{\bibfnamefont{K.}~\bibnamefont{Chang}},
  \bibnamefont{et~al.}, \bibinfo{journal}{Chinese Phys. Lett.}
  \textbf{\bibinfo{volume}{29}}, \bibinfo{pages}{037402}
  (\bibinfo{year}{2012}).

\bibitem[{\citenamefont{He et~al.}(2013)\citenamefont{He, He, Zhang, Zhao, Liu,
  Liu, Mou, Ou, Wang, Li et~al.}}]{He2013}
\bibinfo{author}{\bibfnamefont{S.}~\bibnamefont{He}},
  \bibinfo{author}{\bibfnamefont{J.}~\bibnamefont{He}},
  \bibinfo{author}{\bibfnamefont{W.}~\bibnamefont{Zhang}},
  \bibinfo{author}{\bibfnamefont{L.}~\bibnamefont{Zhao}},
  \bibinfo{author}{\bibfnamefont{D.}~\bibnamefont{Liu}},
  \bibinfo{author}{\bibfnamefont{X.}~\bibnamefont{Liu}},
  \bibinfo{author}{\bibfnamefont{D.}~\bibnamefont{Mou}},
  \bibinfo{author}{\bibfnamefont{Y.-B.} \bibnamefont{Ou}},
  \bibinfo{author}{\bibfnamefont{Q.-Y.} \bibnamefont{Wang}},
  \bibinfo{author}{\bibfnamefont{Z.}~\bibnamefont{Li}}, \bibnamefont{et~al.},
  \bibinfo{journal}{Nat. Mater.} \textbf{\bibinfo{volume}{12}},
  \bibinfo{pages}{605} (\bibinfo{year}{2013}).

\bibitem[{pla()}]{placeholder}
\bibinfo{journal}{See Supplemental Material, which includes Refs. [44-59]}
  (????).

\bibitem[{\citenamefont{Allen and Dynes}(1975)}]{allendynes}
\bibinfo{author}{\bibfnamefont{P.~B.} \bibnamefont{Allen}} \bibnamefont{and}
  \bibinfo{author}{\bibfnamefont{R.~C.} \bibnamefont{Dynes}},
  \bibinfo{journal}{Phys. Rev. B} \textbf{\bibinfo{volume}{12}},
  \bibinfo{pages}{905} (\bibinfo{year}{1975}).

\bibitem[{\citenamefont{Lu et~al.}({2012})\citenamefont{Lu, Vishik, Yi, Chen,
  Moore, and Shen}}]{vishik}
\bibinfo{author}{\bibfnamefont{D.}~\bibnamefont{Lu}},
  \bibinfo{author}{\bibfnamefont{I.~M.} \bibnamefont{Vishik}},
  \bibinfo{author}{\bibfnamefont{M.}~\bibnamefont{Yi}},
  \bibinfo{author}{\bibfnamefont{Y.}~\bibnamefont{Chen}},
  \bibinfo{author}{\bibfnamefont{R.~G.} \bibnamefont{Moore}}, \bibnamefont{and}
  \bibinfo{author}{\bibfnamefont{Z.-X.} \bibnamefont{Shen}}, in
  \emph{\bibinfo{booktitle}{{ANNUAL REVIEW OF CONDENSED MATTER PHYSICS, VOL
  3}}}, edited by \bibinfo{editor}{\bibnamefont{{Langer, JS}}}
  (\bibinfo{publisher}{{ANNUAL REVIEWS}}, \bibinfo{address}{{Palo Alto}},
  \bibinfo{year}{{2012}}), vol.~\bibinfo{volume}{{3}} of
  \emph{\bibinfo{series}{{Annual Review of Condensed Matter Physics}}}, pp.
  \bibinfo{pages}{{129--167}}.

\bibitem[{\citenamefont{He et~al.}(2008)\citenamefont{He, Tanaka, Mo, Sasagawa,
  Fujita, Adachi, Mannella, Yamada, Koike, Hussain et~al.}}]{heLBCO}
\bibinfo{author}{\bibfnamefont{R.-H.} \bibnamefont{He}},
  \bibinfo{author}{\bibfnamefont{K.}~\bibnamefont{Tanaka}},
  \bibinfo{author}{\bibfnamefont{S.-K.} \bibnamefont{Mo}},
  \bibinfo{author}{\bibfnamefont{T.}~\bibnamefont{Sasagawa}},
  \bibinfo{author}{\bibfnamefont{M.}~\bibnamefont{Fujita}},
  \bibinfo{author}{\bibfnamefont{T.}~\bibnamefont{Adachi}},
  \bibinfo{author}{\bibfnamefont{N.}~\bibnamefont{Mannella}},
  \bibinfo{author}{\bibfnamefont{K.}~\bibnamefont{Yamada}},
  \bibinfo{author}{\bibfnamefont{Y.}~\bibnamefont{Koike}},
  \bibinfo{author}{\bibfnamefont{Z.}~\bibnamefont{Hussain}},
  \bibnamefont{et~al.}, \bibinfo{journal}{Nat. Phys.}
  \textbf{\bibinfo{volume}{5}}, \bibinfo{pages}{119} (\bibinfo{year}{2008}).

\bibitem[{\citenamefont{{Hashimoto} et~al.}(2014)\citenamefont{{Hashimoto},
  {Nowadnick}, {He}, {Vishik}, {Moritz}, {He}, {Tanaka}, {Moore}, {Lu},
  {Yoshida} et~al.}}]{newvishik}
\bibinfo{author}{\bibfnamefont{M.}~\bibnamefont{{Hashimoto}}},
  \bibinfo{author}{\bibfnamefont{E.~A.} \bibnamefont{{Nowadnick}}},
  \bibinfo{author}{\bibfnamefont{R.-H.} \bibnamefont{{He}}},
  \bibinfo{author}{\bibfnamefont{I.~M.} \bibnamefont{{Vishik}}},
  \bibinfo{author}{\bibfnamefont{B.}~\bibnamefont{{Moritz}}},
  \bibinfo{author}{\bibfnamefont{Y.}~\bibnamefont{{He}}},
  \bibinfo{author}{\bibfnamefont{K.}~\bibnamefont{{Tanaka}}},
  \bibinfo{author}{\bibfnamefont{R.~G.} \bibnamefont{{Moore}}},
  \bibinfo{author}{\bibfnamefont{D.}~\bibnamefont{{Lu}}},
  \bibinfo{author}{\bibfnamefont{Y.}~\bibnamefont{{Yoshida}}},
  \bibnamefont{et~al.}, \bibinfo{journal}{ArXiv e-prints}
  (\bibinfo{year}{2014}), \eprint{1405.5199}.

\bibitem[{\citenamefont{{Blanco-Canosa}
  et~al.}(2013)\citenamefont{{Blanco-Canosa}, {Frano}, {Loew}, {Lu}, {Porras},
  {Ghiringhelli}, {Minola}, {Mazzoli}, {Braicovich}, {Schierle}
  et~al.}}]{keimer2013}
\bibinfo{author}{\bibfnamefont{S.}~\bibnamefont{{Blanco-Canosa}}},
  \bibinfo{author}{\bibfnamefont{A.}~\bibnamefont{{Frano}}},
  \bibinfo{author}{\bibfnamefont{T.}~\bibnamefont{{Loew}}},
  \bibinfo{author}{\bibfnamefont{Y.}~\bibnamefont{{Lu}}},
  \bibinfo{author}{\bibfnamefont{J.}~\bibnamefont{{Porras}}},
  \bibinfo{author}{\bibfnamefont{G.}~\bibnamefont{{Ghiringhelli}}},
  \bibinfo{author}{\bibfnamefont{M.}~\bibnamefont{{Minola}}},
  \bibinfo{author}{\bibfnamefont{C.}~\bibnamefont{{Mazzoli}}},
  \bibinfo{author}{\bibfnamefont{L.}~\bibnamefont{{Braicovich}}},
  \bibinfo{author}{\bibfnamefont{E.}~\bibnamefont{{Schierle}}},
  \bibnamefont{et~al.}, \bibinfo{journal}{Physical Review Letters}
  \textbf{\bibinfo{volume}{110}}, \bibinfo{eid}{187001} (\bibinfo{year}{2013}),
  \eprint{1212.5580}.

\bibitem[{\citenamefont{{Thampy} et~al.}(2013)\citenamefont{{Thampy},
  {Blanco-Canosa}, {Garc{\'{\i}}a-Fern{\'a}ndez}, {Dean}, {Gu}, {F{\"o}rst},
  {Loew}, {Keimer}, {Le Tacon}, {Wilkins} et~al.}}]{hill2013}
\bibinfo{author}{\bibfnamefont{V.}~\bibnamefont{{Thampy}}},
  \bibinfo{author}{\bibfnamefont{S.}~\bibnamefont{{Blanco-Canosa}}},
  \bibinfo{author}{\bibfnamefont{M.}~\bibnamefont{{Garc{\'{\i}}a-Fern{\'a}ndez}}},
  \bibinfo{author}{\bibfnamefont{M.~P.~M.} \bibnamefont{{Dean}}},
  \bibinfo{author}{\bibfnamefont{G.~D.} \bibnamefont{{Gu}}},
  \bibinfo{author}{\bibfnamefont{M.}~\bibnamefont{{F{\"o}rst}}},
  \bibinfo{author}{\bibfnamefont{T.}~\bibnamefont{{Loew}}},
  \bibinfo{author}{\bibfnamefont{B.}~\bibnamefont{{Keimer}}},
  \bibinfo{author}{\bibfnamefont{M.}~\bibnamefont{{Le Tacon}}},
  \bibinfo{author}{\bibfnamefont{S.~B.} \bibnamefont{{Wilkins}}},
  \bibnamefont{et~al.}, \bibinfo{journal}{\prb} \textbf{\bibinfo{volume}{88}},
  \bibinfo{eid}{024505} (\bibinfo{year}{2013}), \eprint{1305.5515}.

\bibitem[{\citenamefont{{Huecker} et~al.}(2014)\citenamefont{{Huecker},
  {Christensen}, {Holmes}, {Blackburn}, {Forgan}, {Liang}, {Bonn}, {Hardy},
  {Gutowski}, {Zimmermann} et~al.}}]{huecker2014}
\bibinfo{author}{\bibfnamefont{M.}~\bibnamefont{{Huecker}}},
  \bibinfo{author}{\bibfnamefont{N.~B.} \bibnamefont{{Christensen}}},
  \bibinfo{author}{\bibfnamefont{A.~T.} \bibnamefont{{Holmes}}},
  \bibinfo{author}{\bibfnamefont{E.}~\bibnamefont{{Blackburn}}},
  \bibinfo{author}{\bibfnamefont{E.~M.} \bibnamefont{{Forgan}}},
  \bibinfo{author}{\bibfnamefont{R.}~\bibnamefont{{Liang}}},
  \bibinfo{author}{\bibfnamefont{D.~A.} \bibnamefont{{Bonn}}},
  \bibinfo{author}{\bibfnamefont{W.~N.} \bibnamefont{{Hardy}}},
  \bibinfo{author}{\bibfnamefont{O.}~\bibnamefont{{Gutowski}}},
  \bibinfo{author}{\bibfnamefont{M.~v.} \bibnamefont{{Zimmermann}}},
  \bibnamefont{et~al.}, \bibinfo{journal}{ArXiv e-prints}
  (\bibinfo{year}{2014}), \eprint{1405.7001}.

\bibitem[{\citenamefont{{da Silva Neto} et~al.}(2014)\citenamefont{{da Silva
  Neto}, {Aynajian}, {Frano}, {Comin}, {Schierle}, {Weschke}, {Gyenis}, {Wen},
  {Schneeloch}, {Xu} et~al.}}]{yazdani2014}
\bibinfo{author}{\bibfnamefont{E.~H.} \bibnamefont{{da Silva Neto}}},
  \bibinfo{author}{\bibfnamefont{P.}~\bibnamefont{{Aynajian}}},
  \bibinfo{author}{\bibfnamefont{A.}~\bibnamefont{{Frano}}},
  \bibinfo{author}{\bibfnamefont{R.}~\bibnamefont{{Comin}}},
  \bibinfo{author}{\bibfnamefont{E.}~\bibnamefont{{Schierle}}},
  \bibinfo{author}{\bibfnamefont{E.}~\bibnamefont{{Weschke}}},
  \bibinfo{author}{\bibfnamefont{A.}~\bibnamefont{{Gyenis}}},
  \bibinfo{author}{\bibfnamefont{J.}~\bibnamefont{{Wen}}},
  \bibinfo{author}{\bibfnamefont{J.}~\bibnamefont{{Schneeloch}}},
  \bibinfo{author}{\bibfnamefont{Z.}~\bibnamefont{{Xu}}}, \bibnamefont{et~al.},
  \bibinfo{journal}{Science} \textbf{\bibinfo{volume}{343}},
  \bibinfo{pages}{393} (\bibinfo{year}{2014}), \eprint{1312.1347}.

\bibitem[{\citenamefont{{Comin} et~al.}(2014)\citenamefont{{Comin}, {Frano},
  {Yee}, {Yoshida}, {Eisaki}, {Schierle}, {Weschke}, {Sutarto}, {He},
  {Soumyanarayanan} et~al.}}]{comin2014}
\bibinfo{author}{\bibfnamefont{R.}~\bibnamefont{{Comin}}},
  \bibinfo{author}{\bibfnamefont{A.}~\bibnamefont{{Frano}}},
  \bibinfo{author}{\bibfnamefont{M.~M.} \bibnamefont{{Yee}}},
  \bibinfo{author}{\bibfnamefont{Y.}~\bibnamefont{{Yoshida}}},
  \bibinfo{author}{\bibfnamefont{H.}~\bibnamefont{{Eisaki}}},
  \bibinfo{author}{\bibfnamefont{E.}~\bibnamefont{{Schierle}}},
  \bibinfo{author}{\bibfnamefont{E.}~\bibnamefont{{Weschke}}},
  \bibinfo{author}{\bibfnamefont{R.}~\bibnamefont{{Sutarto}}},
  \bibinfo{author}{\bibfnamefont{F.}~\bibnamefont{{He}}},
  \bibinfo{author}{\bibfnamefont{A.}~\bibnamefont{{Soumyanarayanan}}},
  \bibnamefont{et~al.}, \bibinfo{journal}{Science}
  \textbf{\bibinfo{volume}{343}}, \bibinfo{pages}{390} (\bibinfo{year}{2014}),
  \eprint{1312.1343}.

\bibitem[{\citenamefont{Vishik et~al.}(2012)\citenamefont{Vishik, Hashimoto,
  He, Lee, Schmitt, Lu, Moore, Zhang, Meevasana, Sasagawa
  et~al.}}]{vishiktrisected}
\bibinfo{author}{\bibfnamefont{I.~M.} \bibnamefont{Vishik}},
  \bibinfo{author}{\bibfnamefont{M.}~\bibnamefont{Hashimoto}},
  \bibinfo{author}{\bibfnamefont{R.-H.} \bibnamefont{He}},
  \bibinfo{author}{\bibfnamefont{W.-S.} \bibnamefont{Lee}},
  \bibinfo{author}{\bibfnamefont{F.}~\bibnamefont{Schmitt}},
  \bibinfo{author}{\bibfnamefont{D.}~\bibnamefont{Lu}},
  \bibinfo{author}{\bibfnamefont{R.~G.} \bibnamefont{Moore}},
  \bibinfo{author}{\bibfnamefont{C.}~\bibnamefont{Zhang}},
  \bibinfo{author}{\bibfnamefont{W.}~\bibnamefont{Meevasana}},
  \bibinfo{author}{\bibfnamefont{T.}~\bibnamefont{Sasagawa}},
  \bibnamefont{et~al.}, \bibinfo{journal}{Proceedings of the National Academy
  of Sciences} \textbf{\bibinfo{volume}{109}}, \bibinfo{pages}{18332}
  (\bibinfo{year}{2012}).

\bibitem[{\citenamefont{Kanigel et~al.}(2006)\citenamefont{Kanigel, Norman,
  Randeria, Chatterjee, Souma, Kaminski, Fretwell, Rosenkranz, Shi, Sato
  et~al.}}]{campuzano}
\bibinfo{author}{\bibfnamefont{A.}~\bibnamefont{Kanigel}},
  \bibinfo{author}{\bibfnamefont{M.~R.} \bibnamefont{Norman}},
  \bibinfo{author}{\bibfnamefont{M.}~\bibnamefont{Randeria}},
  \bibinfo{author}{\bibfnamefont{U.}~\bibnamefont{Chatterjee}},
  \bibinfo{author}{\bibfnamefont{S.}~\bibnamefont{Souma}},
  \bibinfo{author}{\bibfnamefont{A.}~\bibnamefont{Kaminski}},
  \bibinfo{author}{\bibfnamefont{H.~M.} \bibnamefont{Fretwell}},
  \bibinfo{author}{\bibfnamefont{S.}~\bibnamefont{Rosenkranz}},
  \bibinfo{author}{\bibfnamefont{M.}~\bibnamefont{Shi}},
  \bibinfo{author}{\bibfnamefont{T.}~\bibnamefont{Sato}}, \bibnamefont{et~al.},
  \bibinfo{journal}{Nat. Phys.} \textbf{\bibinfo{volume}{2}},
  \bibinfo{pages}{447} (\bibinfo{year}{2006}).

\bibitem[{\citenamefont{Inosov et~al.}(2010)\citenamefont{Inosov, Park,
  Bourges, Sun, Sidis, Schneidewind, Hradil, Haug, Lin, Keimer
  et~al.}}]{Inosov2010}
\bibinfo{author}{\bibfnamefont{D.~S.} \bibnamefont{Inosov}},
  \bibinfo{author}{\bibfnamefont{J.~T.} \bibnamefont{Park}},
  \bibinfo{author}{\bibfnamefont{P.}~\bibnamefont{Bourges}},
  \bibinfo{author}{\bibfnamefont{D.~L.} \bibnamefont{Sun}},
  \bibinfo{author}{\bibfnamefont{Y.}~\bibnamefont{Sidis}},
  \bibinfo{author}{\bibfnamefont{A.}~\bibnamefont{Schneidewind}},
  \bibinfo{author}{\bibfnamefont{K.}~\bibnamefont{Hradil}},
  \bibinfo{author}{\bibfnamefont{D.}~\bibnamefont{Haug}},
  \bibinfo{author}{\bibfnamefont{C.~T.} \bibnamefont{Lin}},
  \bibinfo{author}{\bibfnamefont{B.}~\bibnamefont{Keimer}},
  \bibnamefont{et~al.}, \bibinfo{journal}{Nature Physics}
  \textbf{\bibinfo{volume}{6}}, \bibinfo{pages}{178} (\bibinfo{year}{2010}).

\bibitem[{\citenamefont{Ning et~al.}(2010)\citenamefont{Ning, Ahilan, Imai,
  Sefat, McGuire, Sales, Mandrus, Cheng, Shen, and Wen}}]{ning2010}
\bibinfo{author}{\bibfnamefont{F.~L.} \bibnamefont{Ning}},
  \bibinfo{author}{\bibfnamefont{K.}~\bibnamefont{Ahilan}},
  \bibinfo{author}{\bibfnamefont{T.}~\bibnamefont{Imai}},
  \bibinfo{author}{\bibfnamefont{A.~S.} \bibnamefont{Sefat}},
  \bibinfo{author}{\bibfnamefont{M.~A.} \bibnamefont{McGuire}},
  \bibinfo{author}{\bibfnamefont{B.~C.} \bibnamefont{Sales}},
  \bibinfo{author}{\bibfnamefont{D.}~\bibnamefont{Mandrus}},
  \bibinfo{author}{\bibfnamefont{P.}~\bibnamefont{Cheng}},
  \bibinfo{author}{\bibfnamefont{B.}~\bibnamefont{Shen}}, \bibnamefont{and}
  \bibinfo{author}{\bibfnamefont{H.-H.} \bibnamefont{Wen}},
  \bibinfo{journal}{Phys. Rev. Lett.} \textbf{\bibinfo{volume}{104}},
  \bibinfo{pages}{037001} (\bibinfo{year}{2010}).

\bibitem[{\citenamefont{Stewart}(2011)}]{stewart}
\bibinfo{author}{\bibfnamefont{G.~R.} \bibnamefont{Stewart}},
  \bibinfo{journal}{Rev. Mod. Phys.} \textbf{\bibinfo{volume}{83}},
  \bibinfo{pages}{1589} (\bibinfo{year}{2011}).

\bibitem[{\citenamefont{Dong et~al.}(2013)\citenamefont{Dong, Wang, and
  Fang}}]{Dong2013}
\bibinfo{author}{\bibfnamefont{C.}~\bibnamefont{Dong}},
  \bibinfo{author}{\bibfnamefont{H.}~\bibnamefont{Wang}}, \bibnamefont{and}
  \bibinfo{author}{\bibfnamefont{M.}~\bibnamefont{Fang}},
  \bibinfo{journal}{Chinese Physics B.} \textbf{\bibinfo{volume}{22}},
  \bibinfo{pages}{087401} (\bibinfo{year}{2013}).

\bibitem[{\citenamefont{Hirschfeld et~al.}(2011)\citenamefont{Hirschfeld,
  Korshunov, and Mazin}}]{Hirschfeld2011}
\bibinfo{author}{\bibfnamefont{P.~J.} \bibnamefont{Hirschfeld}},
  \bibinfo{author}{\bibfnamefont{M.~M.} \bibnamefont{Korshunov}},
  \bibnamefont{and} \bibinfo{author}{\bibfnamefont{I.~I.} \bibnamefont{Mazin}},
  \bibinfo{journal}{Reports on Progress in Physics}
  \textbf{\bibinfo{volume}{74}}, \bibinfo{pages}{124508}
  (\bibinfo{year}{2011}).

\end{thebibliography}


\begin{thebibliography}{27}
\expandafter\ifx\csname natexlab\endcsname\relax\def\natexlab#1{#1}\fi
\expandafter\ifx\csname bibnamefont\endcsname\relax
  \def\bibnamefont#1{#1}\fi
\expandafter\ifx\csname bibfnamefont\endcsname\relax
  \def\bibfnamefont#1{#1}\fi
\expandafter\ifx\csname citenamefont\endcsname\relax
  \def\citenamefont#1{#1}\fi
\expandafter\ifx\csname url\endcsname\relax
  \def\url#1{\texttt{#1}}\fi
\expandafter\ifx\csname urlprefix\endcsname\relax\def\urlprefix{URL }\fi
\providecommand{\bibinfo}[2]{#2}
\providecommand{\eprint}[2][]{\url{#2}}

\bibitem[{\citenamefont{Allen and Dynes}(1975)}]{allendynes}
\bibinfo{author}{\bibfnamefont{P.~B.} \bibnamefont{Allen}} \bibnamefont{and}
  \bibinfo{author}{\bibfnamefont{R.~C.} \bibnamefont{Dynes}},
  \bibinfo{journal}{Phys. Rev. B} \textbf{\bibinfo{volume}{12}},
  \bibinfo{pages}{905} (\bibinfo{year}{1975}).

\bibitem[{\citenamefont{Ramshaw}(2014)}]{ramshaw2014}
\bibinfo{author}{\bibfnamefont{B.}~\bibnamefont{Ramshaw}}, \bibinfo{journal}{et
  al, unpublished manuscript}  (\bibinfo{year}{2014}).

\bibitem[{\citenamefont{Grissonnanche et~al.}(2014)\citenamefont{Grissonnanche,
  Cyr-Choinire, LalibertŽ, RenŽde~Cotret, Juneau-Fecteau, Dufour-BeausŽjour,
  Delage, LeBoeuf, Chang, Ramshaw et~al.}}]{grissonnanche2014}
\bibinfo{author}{\bibfnamefont{G.}~\bibnamefont{Grissonnanche}},
  \bibinfo{author}{\bibfnamefont{O.}~\bibnamefont{Cyr-Choinire}},
  \bibinfo{author}{\bibfnamefont{F.}~\bibnamefont{LalibertŽ}},
  \bibinfo{author}{\bibfnamefont{S.}~\bibnamefont{RenŽde~Cotret}},
  \bibinfo{author}{\bibfnamefont{A.}~\bibnamefont{Juneau-Fecteau}},
  \bibinfo{author}{\bibfnamefont{S.}~\bibnamefont{Dufour-BeausŽjour}},
  \bibinfo{author}{\bibfnamefont{M.~é.} \bibnamefont{Delage}},
  \bibinfo{author}{\bibfnamefont{D.}~\bibnamefont{LeBoeuf}},
  \bibinfo{author}{\bibfnamefont{J.}~\bibnamefont{Chang}},
  \bibinfo{author}{\bibfnamefont{B.~J.} \bibnamefont{Ramshaw}},
  \bibnamefont{et~al.}, \bibinfo{journal}{Nat Commun}
  \textbf{\bibinfo{volume}{5}} (\bibinfo{year}{2014}).

\bibitem[{\citenamefont{Lu et~al.}({2012})\citenamefont{Lu, Vishik, Yi, Chen,
  Moore, and Shen}}]{vishik}
\bibinfo{author}{\bibfnamefont{D.}~\bibnamefont{Lu}},
  \bibinfo{author}{\bibfnamefont{I.~M.} \bibnamefont{Vishik}},
  \bibinfo{author}{\bibfnamefont{M.}~\bibnamefont{Yi}},
  \bibinfo{author}{\bibfnamefont{Y.}~\bibnamefont{Chen}},
  \bibinfo{author}{\bibfnamefont{R.~G.} \bibnamefont{Moore}}, \bibnamefont{and}
  \bibinfo{author}{\bibfnamefont{Z.-X.} \bibnamefont{Shen}}, in
  \emph{\bibinfo{booktitle}{{Annual Review of Condensed Matter Physics}}},
  edited by \bibinfo{editor}{\bibnamefont{{Langer, JS}}}
  (\bibinfo{publisher}{{Annual Reviews}}, \bibinfo{address}{{Palo Alto}},
  \bibinfo{year}{{2012}}), vol.~\bibinfo{volume}{{3}} of
  \emph{\bibinfo{series}{{Annual Review of Condensed Matter Physics}}}, pp.
  \bibinfo{pages}{{129--167}}.

\bibitem[{\citenamefont{He et~al.}(2008)\citenamefont{He, Tanaka, Mo, Sasagawa,
  Fujita, Adachi, Mannella, Yamada, Koike, Hussain et~al.}}]{heLBCO}
\bibinfo{author}{\bibfnamefont{R.-H.} \bibnamefont{He}},
  \bibinfo{author}{\bibfnamefont{K.}~\bibnamefont{Tanaka}},
  \bibinfo{author}{\bibfnamefont{S.-K.} \bibnamefont{Mo}},
  \bibinfo{author}{\bibfnamefont{T.}~\bibnamefont{Sasagawa}},
  \bibinfo{author}{\bibfnamefont{M.}~\bibnamefont{Fujita}},
  \bibinfo{author}{\bibfnamefont{T.}~\bibnamefont{Adachi}},
  \bibinfo{author}{\bibfnamefont{N.}~\bibnamefont{Mannella}},
  \bibinfo{author}{\bibfnamefont{K.}~\bibnamefont{Yamada}},
  \bibinfo{author}{\bibfnamefont{Y.}~\bibnamefont{Koike}},
  \bibinfo{author}{\bibfnamefont{Z.}~\bibnamefont{Hussain}},
  \bibnamefont{et~al.}, \bibinfo{journal}{Nat. Phys.}
  \textbf{\bibinfo{volume}{5}}, \bibinfo{pages}{119} (\bibinfo{year}{2008}).

\bibitem[{\citenamefont{{Hashimoto} et~al.}(2014)\citenamefont{{Hashimoto},
  {Nowadnick}, {He}, {Vishik}, {Moritz}, {He}, {Tanaka}, {Moore}, {Lu},
  {Yoshida} et~al.}}]{newvishik}
\bibinfo{author}{\bibfnamefont{M.}~\bibnamefont{{Hashimoto}}},
  \bibinfo{author}{\bibfnamefont{E.~A.} \bibnamefont{{Nowadnick}}},
  \bibinfo{author}{\bibfnamefont{R.-H.} \bibnamefont{{He}}},
  \bibinfo{author}{\bibfnamefont{I.~M.} \bibnamefont{{Vishik}}},
  \bibinfo{author}{\bibfnamefont{B.}~\bibnamefont{{Moritz}}},
  \bibinfo{author}{\bibfnamefont{Y.}~\bibnamefont{{He}}},
  \bibinfo{author}{\bibfnamefont{K.}~\bibnamefont{{Tanaka}}},
  \bibinfo{author}{\bibfnamefont{R.~G.} \bibnamefont{{Moore}}},
  \bibinfo{author}{\bibfnamefont{D.}~\bibnamefont{{Lu}}},
  \bibinfo{author}{\bibfnamefont{Y.}~\bibnamefont{{Yoshida}}},
  \bibnamefont{et~al.}, \bibinfo{journal}{ArXiv e-prints}
  (\bibinfo{year}{2014}), \eprint{1405.5199}.

\bibitem[{\citenamefont{{Blanco-Canosa}
  et~al.}(2013)\citenamefont{{Blanco-Canosa}, {Frano}, {Loew}, {Lu}, {Porras},
  {Ghiringhelli}, {Minola}, {Mazzoli}, {Braicovich}, {Schierle}
  et~al.}}]{keimer2013}
\bibinfo{author}{\bibfnamefont{S.}~\bibnamefont{{Blanco-Canosa}}},
  \bibinfo{author}{\bibfnamefont{A.}~\bibnamefont{{Frano}}},
  \bibinfo{author}{\bibfnamefont{T.}~\bibnamefont{{Loew}}},
  \bibinfo{author}{\bibfnamefont{Y.}~\bibnamefont{{Lu}}},
  \bibinfo{author}{\bibfnamefont{J.}~\bibnamefont{{Porras}}},
  \bibinfo{author}{\bibfnamefont{G.}~\bibnamefont{{Ghiringhelli}}},
  \bibinfo{author}{\bibfnamefont{M.}~\bibnamefont{{Minola}}},
  \bibinfo{author}{\bibfnamefont{C.}~\bibnamefont{{Mazzoli}}},
  \bibinfo{author}{\bibfnamefont{L.}~\bibnamefont{{Braicovich}}},
  \bibinfo{author}{\bibfnamefont{E.}~\bibnamefont{{Schierle}}},
  \bibnamefont{et~al.}, \bibinfo{journal}{Physical Review Letters}
  \textbf{\bibinfo{volume}{110}}, \bibinfo{eid}{187001} (\bibinfo{year}{2013}),
  \eprint{1212.5580}.

\bibitem[{\citenamefont{{Thampy} et~al.}(2013)\citenamefont{{Thampy},
  {Blanco-Canosa}, {Garc{\'{\i}}a-Fern{\'a}ndez}, {Dean}, {Gu}, {F{\"o}rst},
  {Loew}, {Keimer}, {Le Tacon}, {Wilkins} et~al.}}]{hill2013}
\bibinfo{author}{\bibfnamefont{V.}~\bibnamefont{{Thampy}}},
  \bibinfo{author}{\bibfnamefont{S.}~\bibnamefont{{Blanco-Canosa}}},
  \bibinfo{author}{\bibfnamefont{M.}~\bibnamefont{{Garc{\'{\i}}a-Fern{\'a}ndez}}},
  \bibinfo{author}{\bibfnamefont{M.~P.~M.} \bibnamefont{{Dean}}},
  \bibinfo{author}{\bibfnamefont{G.~D.} \bibnamefont{{Gu}}},
  \bibinfo{author}{\bibfnamefont{M.}~\bibnamefont{{F{\"o}rst}}},
  \bibinfo{author}{\bibfnamefont{T.}~\bibnamefont{{Loew}}},
  \bibinfo{author}{\bibfnamefont{B.}~\bibnamefont{{Keimer}}},
  \bibinfo{author}{\bibfnamefont{M.}~\bibnamefont{{Le Tacon}}},
  \bibinfo{author}{\bibfnamefont{S.~B.} \bibnamefont{{Wilkins}}},
  \bibnamefont{et~al.}, \bibinfo{journal}{\prb} \textbf{\bibinfo{volume}{88}},
  \bibinfo{eid}{024505} (\bibinfo{year}{2013}), \eprint{1305.5515}.

\bibitem[{\citenamefont{{Huecker} et~al.}(2014)\citenamefont{{Huecker},
  {Christensen}, {Holmes}, {Blackburn}, {Forgan}, {Liang}, {Bonn}, {Hardy},
  {Gutowski}, {Zimmermann} et~al.}}]{huecker2014}
\bibinfo{author}{\bibfnamefont{M.}~\bibnamefont{{Huecker}}},
  \bibinfo{author}{\bibfnamefont{N.~B.} \bibnamefont{{Christensen}}},
  \bibinfo{author}{\bibfnamefont{A.~T.} \bibnamefont{{Holmes}}},
  \bibinfo{author}{\bibfnamefont{E.}~\bibnamefont{{Blackburn}}},
  \bibinfo{author}{\bibfnamefont{E.~M.} \bibnamefont{{Forgan}}},
  \bibinfo{author}{\bibfnamefont{R.}~\bibnamefont{{Liang}}},
  \bibinfo{author}{\bibfnamefont{D.~A.} \bibnamefont{{Bonn}}},
  \bibinfo{author}{\bibfnamefont{W.~N.} \bibnamefont{{Hardy}}},
  \bibinfo{author}{\bibfnamefont{O.}~\bibnamefont{{Gutowski}}},
  \bibinfo{author}{\bibfnamefont{M.~v.} \bibnamefont{{Zimmermann}}},
  \bibnamefont{et~al.}, \bibinfo{journal}{ArXiv e-prints}
  (\bibinfo{year}{2014}), \eprint{1405.7001}.

\bibitem[{\citenamefont{{da Silva Neto} et~al.}(2014)\citenamefont{{da Silva
  Neto}, {Aynajian}, {Frano}, {Comin}, {Schierle}, {Weschke}, {Gyenis}, {Wen},
  {Schneeloch}, {Xu} et~al.}}]{yazdani2014}
\bibinfo{author}{\bibfnamefont{E.~H.} \bibnamefont{{da Silva Neto}}},
  \bibinfo{author}{\bibfnamefont{P.}~\bibnamefont{{Aynajian}}},
  \bibinfo{author}{\bibfnamefont{A.}~\bibnamefont{{Frano}}},
  \bibinfo{author}{\bibfnamefont{R.}~\bibnamefont{{Comin}}},
  \bibinfo{author}{\bibfnamefont{E.}~\bibnamefont{{Schierle}}},
  \bibinfo{author}{\bibfnamefont{E.}~\bibnamefont{{Weschke}}},
  \bibinfo{author}{\bibfnamefont{A.}~\bibnamefont{{Gyenis}}},
  \bibinfo{author}{\bibfnamefont{J.}~\bibnamefont{{Wen}}},
  \bibinfo{author}{\bibfnamefont{J.}~\bibnamefont{{Schneeloch}}},
  \bibinfo{author}{\bibfnamefont{Z.}~\bibnamefont{{Xu}}}, \bibnamefont{et~al.},
  \bibinfo{journal}{Science} \textbf{\bibinfo{volume}{343}},
  \bibinfo{pages}{393} (\bibinfo{year}{2014}), \eprint{1312.1347}.

\bibitem[{\citenamefont{{Comin} et~al.}(2014)\citenamefont{{Comin}, {Frano},
  {Yee}, {Yoshida}, {Eisaki}, {Schierle}, {Weschke}, {Sutarto}, {He},
  {Soumyanarayanan} et~al.}}]{comin2014}
\bibinfo{author}{\bibfnamefont{R.}~\bibnamefont{{Comin}}},
  \bibinfo{author}{\bibfnamefont{A.}~\bibnamefont{{Frano}}},
  \bibinfo{author}{\bibfnamefont{M.~M.} \bibnamefont{{Yee}}},
  \bibinfo{author}{\bibfnamefont{Y.}~\bibnamefont{{Yoshida}}},
  \bibinfo{author}{\bibfnamefont{H.}~\bibnamefont{{Eisaki}}},
  \bibinfo{author}{\bibfnamefont{E.}~\bibnamefont{{Schierle}}},
  \bibinfo{author}{\bibfnamefont{E.}~\bibnamefont{{Weschke}}},
  \bibinfo{author}{\bibfnamefont{R.}~\bibnamefont{{Sutarto}}},
  \bibinfo{author}{\bibfnamefont{F.}~\bibnamefont{{He}}},
  \bibinfo{author}{\bibfnamefont{A.}~\bibnamefont{{Soumyanarayanan}}},
  \bibnamefont{et~al.}, \bibinfo{journal}{Science}
  \textbf{\bibinfo{volume}{343}}, \bibinfo{pages}{390} (\bibinfo{year}{2014}),
  \eprint{1312.1343}.

\bibitem[{\citenamefont{Nie et~al.}(2013)\citenamefont{Nie, Tarjus, and
  Kivelson}}]{nie-2013}
\bibinfo{author}{\bibfnamefont{L.}~\bibnamefont{Nie}},
  \bibinfo{author}{\bibfnamefont{G.}~\bibnamefont{Tarjus}}, \bibnamefont{and}
  \bibinfo{author}{\bibfnamefont{S.~A.} \bibnamefont{Kivelson}}
  (\bibinfo{year}{2013}), \bibinfo{note}{unpublished},
  \eprint{arXiv:1311.5580}.

\bibitem[{\citenamefont{Vishik et~al.}(2012)\citenamefont{Vishik, Hashimoto,
  He, Lee, Schmitt, Lu, Moore, Zhang, Meevasana, Sasagawa
  et~al.}}]{vishiktrisected}
\bibinfo{author}{\bibfnamefont{I.~M.} \bibnamefont{Vishik}},
  \bibinfo{author}{\bibfnamefont{M.}~\bibnamefont{Hashimoto}},
  \bibinfo{author}{\bibfnamefont{R.-H.} \bibnamefont{He}},
  \bibinfo{author}{\bibfnamefont{W.-S.} \bibnamefont{Lee}},
  \bibinfo{author}{\bibfnamefont{F.}~\bibnamefont{Schmitt}},
  \bibinfo{author}{\bibfnamefont{D.}~\bibnamefont{Lu}},
  \bibinfo{author}{\bibfnamefont{R.~G.} \bibnamefont{Moore}},
  \bibinfo{author}{\bibfnamefont{C.}~\bibnamefont{Zhang}},
  \bibinfo{author}{\bibfnamefont{W.}~\bibnamefont{Meevasana}},
  \bibinfo{author}{\bibfnamefont{T.}~\bibnamefont{Sasagawa}},
  \bibnamefont{et~al.}, \bibinfo{journal}{Proceedings of the National Academy
  of Sciences} \textbf{\bibinfo{volume}{109}}, \bibinfo{pages}{18332}
  (\bibinfo{year}{2012}).

\bibitem[{\citenamefont{Kanigel et~al.}(2006)\citenamefont{Kanigel, Norman,
  Randeria, Chatterjee, Souma, Kaminski, Fretwell, Rosenkranz, Shi, Sato
  et~al.}}]{campuzano}
\bibinfo{author}{\bibfnamefont{A.}~\bibnamefont{Kanigel}},
  \bibinfo{author}{\bibfnamefont{M.~R.} \bibnamefont{Norman}},
  \bibinfo{author}{\bibfnamefont{M.}~\bibnamefont{Randeria}},
  \bibinfo{author}{\bibfnamefont{U.}~\bibnamefont{Chatterjee}},
  \bibinfo{author}{\bibfnamefont{S.}~\bibnamefont{Souma}},
  \bibinfo{author}{\bibfnamefont{A.}~\bibnamefont{Kaminski}},
  \bibinfo{author}{\bibfnamefont{H.~M.} \bibnamefont{Fretwell}},
  \bibinfo{author}{\bibfnamefont{S.}~\bibnamefont{Rosenkranz}},
  \bibinfo{author}{\bibfnamefont{M.}~\bibnamefont{Shi}},
  \bibinfo{author}{\bibfnamefont{T.}~\bibnamefont{Sato}}, \bibnamefont{et~al.},
  \bibinfo{journal}{Nat. Phys.} \textbf{\bibinfo{volume}{2}},
  \bibinfo{pages}{447} (\bibinfo{year}{2006}).

\bibitem[{\citenamefont{Yang and Sondhi}(2000)}]{kunyang2000}
\bibinfo{author}{\bibfnamefont{K.}~\bibnamefont{Yang}} \bibnamefont{and}
  \bibinfo{author}{\bibfnamefont{S.~L.} \bibnamefont{Sondhi}},
  \bibinfo{journal}{Phys. Rev. B} \textbf{\bibinfo{volume}{62}},
  \bibinfo{pages}{11778} (\bibinfo{year}{2000}).

\bibitem[{\citenamefont{Chu et~al.}(2010)\citenamefont{Chu, Analytis, {De
  Greve}, McMahon, Islam, Yamamoto, and Fisher}}]{fisher-2010}
\bibinfo{author}{\bibfnamefont{J.-H.} \bibnamefont{Chu}},
  \bibinfo{author}{\bibfnamefont{J.~G.} \bibnamefont{Analytis}},
  \bibinfo{author}{\bibfnamefont{K.}~\bibnamefont{{De Greve}}},
  \bibinfo{author}{\bibfnamefont{P.~L.} \bibnamefont{McMahon}},
  \bibinfo{author}{\bibfnamefont{Z.}~\bibnamefont{Islam}},
  \bibinfo{author}{\bibfnamefont{Y.}~\bibnamefont{Yamamoto}}, \bibnamefont{and}
  \bibinfo{author}{\bibfnamefont{I.~R.} \bibnamefont{Fisher}},
  \bibinfo{journal}{Science} \textbf{\bibinfo{volume}{329}},
  \bibinfo{pages}{824} (\bibinfo{year}{2010}).

\bibitem[{\citenamefont{{Chu} et~al.}(2012)\citenamefont{{Chu}, {Kuo},
  {Analytis}, and {Fisher}}}]{chu2012}
\bibinfo{author}{\bibfnamefont{J.-H.} \bibnamefont{{Chu}}},
  \bibinfo{author}{\bibfnamefont{H.-H.} \bibnamefont{{Kuo}}},
  \bibinfo{author}{\bibfnamefont{J.~G.} \bibnamefont{{Analytis}}},
  \bibnamefont{and} \bibinfo{author}{\bibfnamefont{I.~R.}
  \bibnamefont{{Fisher}}}, \bibinfo{journal}{Science}
  \textbf{\bibinfo{volume}{337}}, \bibinfo{pages}{710} (\bibinfo{year}{2012}),
  \eprint{1203.3239}.

\bibitem[{\citenamefont{{Yoshizawa} et~al.}(2012)\citenamefont{{Yoshizawa},
  {Kimura}, {Chiba}, {Simayi}, {Nakanishi}, {Kihou}, {Lee}, {Iyo}, {Eisaki},
  {Nakajima} et~al.}}]{yoshizawa2012}
\bibinfo{author}{\bibfnamefont{M.}~\bibnamefont{{Yoshizawa}}},
  \bibinfo{author}{\bibfnamefont{D.}~\bibnamefont{{Kimura}}},
  \bibinfo{author}{\bibfnamefont{T.}~\bibnamefont{{Chiba}}},
  \bibinfo{author}{\bibfnamefont{S.}~\bibnamefont{{Simayi}}},
  \bibinfo{author}{\bibfnamefont{Y.}~\bibnamefont{{Nakanishi}}},
  \bibinfo{author}{\bibfnamefont{K.}~\bibnamefont{{Kihou}}},
  \bibinfo{author}{\bibfnamefont{C.-H.} \bibnamefont{{Lee}}},
  \bibinfo{author}{\bibfnamefont{A.}~\bibnamefont{{Iyo}}},
  \bibinfo{author}{\bibfnamefont{H.}~\bibnamefont{{Eisaki}}},
  \bibinfo{author}{\bibfnamefont{M.}~\bibnamefont{{Nakajima}}},
  \bibnamefont{et~al.}, \bibinfo{journal}{Journal of the Physical Society of
  Japan} \textbf{\bibinfo{volume}{81}}, \bibinfo{pages}{024604}
  (\bibinfo{year}{2012}), \eprint{1111.0366}.

\bibitem[{\citenamefont{Walmsley et~al.}({2013})\citenamefont{Walmsley, Putzke,
  Malone, Guillamon, Vignolles, Proust, Badoux, Coldea, Watson, Kasahara
  et~al.}}]{matsuda2013}
\bibinfo{author}{\bibfnamefont{P.}~\bibnamefont{Walmsley}},
  \bibinfo{author}{\bibfnamefont{C.}~\bibnamefont{Putzke}},
  \bibinfo{author}{\bibfnamefont{L.}~\bibnamefont{Malone}},
  \bibinfo{author}{\bibfnamefont{I.}~\bibnamefont{Guillamon}},
  \bibinfo{author}{\bibfnamefont{D.}~\bibnamefont{Vignolles}},
  \bibinfo{author}{\bibfnamefont{C.}~\bibnamefont{Proust}},
  \bibinfo{author}{\bibfnamefont{S.}~\bibnamefont{Badoux}},
  \bibinfo{author}{\bibfnamefont{A.~I.} \bibnamefont{Coldea}},
  \bibinfo{author}{\bibfnamefont{M.~D.} \bibnamefont{Watson}},
  \bibinfo{author}{\bibfnamefont{S.}~\bibnamefont{Kasahara}},
  \bibnamefont{et~al.}, \bibinfo{journal}{{PHYSICAL REVIEW LETTERS}}
  \textbf{\bibinfo{volume}{{110}}} (\bibinfo{year}{{2013}}).

\bibitem[{\citenamefont{Inosov et~al.}(2010)\citenamefont{Inosov, Park,
  Bourges, Sun, Sidis, Schneidewind, Hradil, Haug, Lin, Keimer
  et~al.}}]{Inosov2010}
\bibinfo{author}{\bibfnamefont{D.~S.} \bibnamefont{Inosov}},
  \bibinfo{author}{\bibfnamefont{J.~T.} \bibnamefont{Park}},
  \bibinfo{author}{\bibfnamefont{P.}~\bibnamefont{Bourges}},
  \bibinfo{author}{\bibfnamefont{D.~L.} \bibnamefont{Sun}},
  \bibinfo{author}{\bibfnamefont{Y.}~\bibnamefont{Sidis}},
  \bibinfo{author}{\bibfnamefont{A.}~\bibnamefont{Schneidewind}},
  \bibinfo{author}{\bibfnamefont{K.}~\bibnamefont{Hradil}},
  \bibinfo{author}{\bibfnamefont{D.}~\bibnamefont{Haug}},
  \bibinfo{author}{\bibfnamefont{C.~T.} \bibnamefont{Lin}},
  \bibinfo{author}{\bibfnamefont{B.}~\bibnamefont{Keimer}},
  \bibnamefont{et~al.}, \bibinfo{journal}{Nature Physics}
  \textbf{\bibinfo{volume}{6}}, \bibinfo{pages}{178} (\bibinfo{year}{2010}).

\bibitem[{\citenamefont{Ning et~al.}(2010)\citenamefont{Ning, Ahilan, Imai,
  Sefat, McGuire, Sales, Mandrus, Cheng, Shen, and Wen}}]{ning2010}
\bibinfo{author}{\bibfnamefont{F.~L.} \bibnamefont{Ning}},
  \bibinfo{author}{\bibfnamefont{K.}~\bibnamefont{Ahilan}},
  \bibinfo{author}{\bibfnamefont{T.}~\bibnamefont{Imai}},
  \bibinfo{author}{\bibfnamefont{A.~S.} \bibnamefont{Sefat}},
  \bibinfo{author}{\bibfnamefont{M.~A.} \bibnamefont{McGuire}},
  \bibinfo{author}{\bibfnamefont{B.~C.} \bibnamefont{Sales}},
  \bibinfo{author}{\bibfnamefont{D.}~\bibnamefont{Mandrus}},
  \bibinfo{author}{\bibfnamefont{P.}~\bibnamefont{Cheng}},
  \bibinfo{author}{\bibfnamefont{B.}~\bibnamefont{Shen}}, \bibnamefont{and}
  \bibinfo{author}{\bibfnamefont{H.-H.} \bibnamefont{Wen}},
  \bibinfo{journal}{Phys. Rev. Lett.} \textbf{\bibinfo{volume}{104}},
  \bibinfo{pages}{037001} (\bibinfo{year}{2010}).

\bibitem[{\citenamefont{Stewart}(2011)}]{stewart}
\bibinfo{author}{\bibfnamefont{G.~R.} \bibnamefont{Stewart}},
  \bibinfo{journal}{Rev. Mod. Phys.} \textbf{\bibinfo{volume}{83}},
  \bibinfo{pages}{1589} (\bibinfo{year}{2011}).

\bibitem[{\citenamefont{Dong et~al.}(2013)\citenamefont{Dong, Wang, and
  Fang}}]{Dong2013}
\bibinfo{author}{\bibfnamefont{C.}~\bibnamefont{Dong}},
  \bibinfo{author}{\bibfnamefont{H.}~\bibnamefont{Wang}}, \bibnamefont{and}
  \bibinfo{author}{\bibfnamefont{M.}~\bibnamefont{Fang}},
  \bibinfo{journal}{Chinese Physics B.} \textbf{\bibinfo{volume}{22}},
  \bibinfo{pages}{087401} (\bibinfo{year}{2013}).

\bibitem[{\citenamefont{Wang et~al.}(2012)\citenamefont{Wang, Li, Zhang, Zhang,
  Zhang, Li, Ding, Ou, Deng, Chang et~al.}}]{Wang2012}
\bibinfo{author}{\bibfnamefont{Q.-Y.} \bibnamefont{Wang}},
  \bibinfo{author}{\bibfnamefont{Z.}~\bibnamefont{Li}},
  \bibinfo{author}{\bibfnamefont{W.-H.} \bibnamefont{Zhang}},
  \bibinfo{author}{\bibfnamefont{Z.-C.} \bibnamefont{Zhang}},
  \bibinfo{author}{\bibfnamefont{J.-S.} \bibnamefont{Zhang}},
  \bibinfo{author}{\bibfnamefont{W.}~\bibnamefont{Li}},
  \bibinfo{author}{\bibfnamefont{H.}~\bibnamefont{Ding}},
  \bibinfo{author}{\bibfnamefont{Y.-B.} \bibnamefont{Ou}},
  \bibinfo{author}{\bibfnamefont{P.}~\bibnamefont{Deng}},
  \bibinfo{author}{\bibfnamefont{K.}~\bibnamefont{Chang}},
  \bibnamefont{et~al.}, \bibinfo{journal}{Chinese Phys. Lett.}
  \textbf{\bibinfo{volume}{29}}, \bibinfo{pages}{037402}
  (\bibinfo{year}{2012}).

\bibitem[{\citenamefont{{Lee} et~al.}(2013)\citenamefont{{Lee}, {Schmitt},
  {Moore}, {Johnston}, {Cui}, {Li}, {Yi}, {Liu}, {Hashimoto}, {Zhang}
  et~al.}}]{Lee2013}
\bibinfo{author}{\bibfnamefont{J.~J.} \bibnamefont{{Lee}}},
  \bibinfo{author}{\bibfnamefont{F.~T.} \bibnamefont{{Schmitt}}},
  \bibinfo{author}{\bibfnamefont{R.~G.} \bibnamefont{{Moore}}},
  \bibinfo{author}{\bibfnamefont{S.}~\bibnamefont{{Johnston}}},
  \bibinfo{author}{\bibfnamefont{Y.-T.} \bibnamefont{{Cui}}},
  \bibinfo{author}{\bibfnamefont{W.}~\bibnamefont{{Li}}},
  \bibinfo{author}{\bibfnamefont{M.}~\bibnamefont{{Yi}}},
  \bibinfo{author}{\bibfnamefont{Z.~K.} \bibnamefont{{Liu}}},
  \bibinfo{author}{\bibfnamefont{M.}~\bibnamefont{{Hashimoto}}},
  \bibinfo{author}{\bibfnamefont{Y.}~\bibnamefont{{Zhang}}},
  \bibnamefont{et~al.}, \bibinfo{journal}{ArXiv e-prints}
  (\bibinfo{year}{2013}), \eprint{1312.2633}.

\bibitem[{\citenamefont{He et~al.}(2013)\citenamefont{He, He, Zhang, Zhao, Liu,
  Liu, Mou, Ou, Wang, Li et~al.}}]{He2013}
\bibinfo{author}{\bibfnamefont{S.}~\bibnamefont{He}},
  \bibinfo{author}{\bibfnamefont{J.}~\bibnamefont{He}},
  \bibinfo{author}{\bibfnamefont{W.}~\bibnamefont{Zhang}},
  \bibinfo{author}{\bibfnamefont{L.}~\bibnamefont{Zhao}},
  \bibinfo{author}{\bibfnamefont{D.}~\bibnamefont{Liu}},
  \bibinfo{author}{\bibfnamefont{X.}~\bibnamefont{Liu}},
  \bibinfo{author}{\bibfnamefont{D.}~\bibnamefont{Mou}},
  \bibinfo{author}{\bibfnamefont{Y.-B.} \bibnamefont{Ou}},
  \bibinfo{author}{\bibfnamefont{Q.-Y.} \bibnamefont{Wang}},
  \bibinfo{author}{\bibfnamefont{Z.}~\bibnamefont{Li}}, \bibnamefont{et~al.},
  \bibinfo{journal}{Nat. Mater.} \textbf{\bibinfo{volume}{12}},
  \bibinfo{pages}{605} (\bibinfo{year}{2013}).

\bibitem[{\citenamefont{Hirschfeld et~al.}(2011)\citenamefont{Hirschfeld,
  Korshunov, and Mazin}}]{Hirschfeld2011}
\bibinfo{author}{\bibfnamefont{P.~J.} \bibnamefont{Hirschfeld}},
  \bibinfo{author}{\bibfnamefont{M.~M.} \bibnamefont{Korshunov}},
  \bibnamefont{and} \bibinfo{author}{\bibfnamefont{I.~I.} \bibnamefont{Mazin}},
  \bibinfo{journal}{Reports on Progress in Physics}
  \textbf{\bibinfo{volume}{74}}, \bibinfo{pages}{124508}
  (\bibinfo{year}{2011}).

\end{thebibliography}

\end{document}



\widetext
\pagebreak
\begin{center}
\textbf{\large 
Supplemental Materials: \\ Enhancement of superconductivity near a nematic quantum critical point}
\end{center}
\endwidetext
\setcounter{equation}{0}
\setcounter{figure}{0}
\setcounter{table}{0}
\setcounter{page}{1}
\makeatletter
\renewcommand{\theequation}{S\arabic{equation}}
\renewcommand{\thefigure}{S\arabic{figure}}

\section{Relation to 
reality}
In the main body of the paper, we  
applied an asymptotically exact perturbative RG approach
in a weak coupling limit.
In this limit, $T_c$ would be exponentially low and the normal state would be a perfect Fermi liquid.  Even in conventional superconductors, $\lambda$ is typically\cite{allendynes} not far from $\lambda=1$, and in the highly correlated materials of most interest in the present context, the normal state above $T_c$ is not well approximated as a Fermi liquid.  Because the assumed existence of a nematic quantum critical point is the organizing principle for the present work, we have stressed the singular (critical) $\delta x$ dependence of  various effects;  however, the realities of experimental physics -- especially the effects of quenched disorder -- will typically cut off all critical divergences, and in any case most experiments concern values of $\delta x$ that are not very small.  This means that the analytic, non-critical $x$ dependences of parameters, such as $\alpha(x)$ and $\rho(E_F, x)$, are likely to play as large a role in determining trends across the phase diagram as do the critical dependences. 

Thus, the {\it most} we can expect from a comparison between the present theory and experiment is to gain some qualitative insight.  With the understanding that all the caveats implied by the above apply, we now summarize a few of the interesting qualitative results that may be 
relevant to the cuprate and Fe-based superconductors:

1) 
The effect of nematic fluctuations on $T_c$ is always much more dramatic than the effect on the pair wave function.  In the regime of weak enhancement, the fractional change to the pair wave function is parametrically small, of order $(\delta\lambda^{(ind)}/\lambda^*)$, whereas the fractional change in $T_c$, (see Eq. 10) 
\be
\frac{\delta T_c}{T_{c0}}=\exp\left[\frac{\delta \lambda^{ind}}{(\lambda^*)^2}\right]-1 \ ,
\ee
is always large compared to $(\delta\lambda^{(ind)}/\lambda^*)$ and, for $\lambda^* > \delta \lambda^{ind}> [\lambda^*]^2$, is {\it exponentially} larger!  
($T_{c0}$ is the $\alpha=0$ value of $T_c$.) 
In the strong-enhancement regime, the gap function is   primarily determined by the induced interaction, but even here, the $k$ dependence of $\Delta_{\hat k}$ is characterized by $\bar \kappa$ which is large compared to $\kappa = 1/\xi$, and which has a power-law dependence on $\delta x$ which should be compared to the exponential dependence of $T_c$ on $\delta x$.
It is a general feature of weak coupling superconductivity that changes in the interactions are felt much more profoundly in the critical temperature (which has an exponential dependence on the interaction) than in the pair wavefunction.  
Thus, associating  the enhanced\cite{ramshaw2014,grissonnanche2014}  $H_{c2}$   in YBCO for $x \sim 0.18$ with nematic quantum critical fluctuations is not inconsistent with the fairly conventional d-wave form of the gap function measured\cite{vishik} in ARPES on BSCCO samples with similar hole-densities - even assuming that the properties of the two materials can be compared.

2) 
It is tempting to compare the highly structured gap functions we have found,  in which the nematic fluctuations contribute significantly to the pairing, with 
the behavior seen\cite{heLBCO,newvishik} in ARPES studies of the electronic structure of the hole doped cuprate Bi2212.  Deep in the superconducting state, the gap as a function of angle along the Fermi surface is 
 more highly peaked at the antinodal end of the Fermi surface than would be expected from the simplest d-wave gap ($\Delta_{\vec k} \propto [\cos(k_x)-\cos(k_y)]$), with this tendency increasingly apparent with increasing underdoping.  
The doping dependence inferred from our
calculations would suggest the opposite trend if we were to limit ourselves to asymptotic analysis of the regime $|\delta x|\ll 1$. 
 However, the antinodal enhancement in Bi2212 is observed over a relatively large doping range,   $|\delta x|/x \lesssim 1$.  
  Various properties, such as the strength of the observed\cite{keimer2013,hill2013,huecker2014,yazdani2014,comin2014} structures associated with short-range CDW order, increase significantly with underdoping.  
 To the extent that the nematic order is thought of as vestigial charge order,\cite{nie-2013} 
 $\alpha$ could be substantially larger for $x\sim 1/8$ (where charge order is strongest) than it is for $x \approx x_c$. 
  If the increase of $\alpha$ with underdoping were fast enough (it need only overcome the logarithmic decrease of $\lambda^{(ind)}$ with $|\delta x|$), this 
  could yield the correct doping dependence of the gap anisotropy within our model. 
  A theory of this would need to start from a more microscopic starting point, involving the underlying CDW degrees of freedom;  however, even the above sketch suggests that a key role of nematic fluctuations in shaping the superconducting state can be squared with the seemingly contradictory experimentally observed $x$ dependences.

 3)  It is also tempting to view the ``Fermi arcs'' observed in the pseudo-gap phase, as decedents of a superconducting gap which is strongly dominated by nematic quantum critical fluctuations, and hence is highly concentrated in the antinodes.  Here, the observed doping trends\cite{vishiktrisected,campuzano} are closer to naive expectations, in that the width of the arc increases with increasing doping (for $x < x_c$) which, conversely, means that the remaining gap is increasingly peaked in the antinode.  However, to make this identification at more than a heuristic level would require a theory of a broad fluctuational regime, something that is  absent in the weak coupling limit we have analyzed unless the special fluctuations associated with pairing dominated by small momentum transfers are significant, as was conjectured by Yang and Sondhi.\cite{kunyang2000}
 
Of the iron-based high-$T_c$ superconductors, the best studied family is the $122$ pnictides (such as doped BaFe$_2$As$_2$), for which there is substantial evidence\cite{fisher-2010,chu2012,yoshizawa2012,matsuda2013} of a quantum critical point near optimal doping associated with the breaking of point-group symmetry. 
However, the application of our theory to these materials is complicated (beyond the general caveats outlined in the beginning of this section)  by the presence of antiferromagnetic fluctuations with a substantial correlation length\cite{Inosov2010,ning2010}.
Other pnictide families (as well as the chalcogenide families) have not been studied as extensively due to difficulties in growing large crystals. However, some diversity of phase diagrams is indicated, including families where optimal doping is well separated from the antiferromagnetic phase, but near a boundary between orthorhombic and tetragonal phases\cite{stewart,Dong2013}. If the low temperature structural transition were continuous or weakly first order in such a family, it would satisfy the qualitative conditions for the validity of our model, and we 
could optimistically associate the maximal $T_c$ with enhancement due to near-critical nematic fluctuations, as described in this paper.

An exciting recent development in the iron-based superconductors has been the observation of superconductivity in a single layer of FeSe grown on SrTiO$_3$, with a critical temperature 
exceeding that of bulk FeSe\cite{Wang2012}. One 
striking feature of   this material is a band replica observed in ARPES which can only be explained by 
inelastic scattering from a $q=0$ mode\cite{Lee2013}.  Another puzzling feature is the absence of a Fermi surface pocket at the $\Gamma$ point\cite{He2013}, which 
plays a key role in  the generally understood mechanism \cite{Hirschfeld2011} of superconductivity from antiferromagnetic fluctuations in the iron-based superconductors. 
 The presence of 2D, near-critical nematic fluctuations in this material may rationalize both of these features. Forward scattering of electrons is a natural consequence of the peak in the nematic susceptibility at small momentum transfer.  Regarding the mechanism of superconductivity and the enhancement of $T_c$, in our model we find an enhancement in all channels of anomalous strength in $d=2$. 
 Such an anomalous enhancement could lead to a much higher $T_c$ than for the bulk system, even if the mechanism responsible for superconductivity in the bulk were 
 weakened due to the absence of the pocket at $\Gamma$.  
 Note, however, that other explanations for each of these features have been proposed\cite{Lee2013}. 

\section{Eigenstates and eigenvalues of $\Gamma^{(ind)}$ }
\label{sec:width}
The exchange of nematic bosons yields a contribution $\Gamma^{(ind)}(\hat k, \hat k^{\prime})$ to the pairing vertex with anomalous structure.  First of all, it is sharply peaked about $\hat k=\hat k^{\prime}$, with width $\kappa \equiv \xi^{-1}$. Further, $\Gamma^{(ind)}(\hat k, \hat k)$ varies over the Fermi surface, with zeroes where $|\hat k_x| = |\hat k_y|$. For a cuprate-like Fermi surface with tetragonal symmetry, $\Gamma^{(ind)}(\hat k, \hat k)$ has four maxima at positions $\hat k_{opt}$ which are even under reflections in the $(0,1,0)$ and $(1,0,0)$ planes. (Note: for this last statement to apply in $d=3$, we must assume that $\Gamma^{(ind)}(\hat k,\hat k)$ has some dependence on $\hat k_z$, otherwise the maxima would be curves on the Fermi surface rather than points. However, such $k_z$ dependence is generically present in both the nematic form factor $f(\hat k,0)$, and the fermi velocity $v(\hat k)$, so this is the case we treat).

The pair wavefunctions $\phi(\hat k)$ with largest eigenvalues will clearly have maximum magnitude at the points $\hat k_{opt}$. In fact, explicit numerical diagonalization shows that $\phi(\hat k)$ is peaked about these points with width $\tilde \kappa\ll k_F$ .  For the present discussion, we thus focus on the neighborhood of a single Fermi surface position $\hat k_{opt}$, and determine the $\kappa$ dependence of the pairing eigenvalue and $\tilde \kappa$ in the limit $\kappa/k_F \ll 1$.  We begin with the expression for $\Gamma^{(ind)}(\hat k, \hat k)$
\be
\Gamma^{(ind)}(\hat k, \hat k)= -\frac{\alpha^2}{4}\left(\frac{f\left(\hat k-\hat k',\frac{1}{2}(\hat  k+\hat  k^{\prime})\right)^2 }{\sqrt{v_{\hat k}v_{\hat k^\prime}}} \right)\chi(\hat  k-\hat  k^\prime,0) .
\ee
If we use the scaling form of the susceptibility $\chi(\hat k - \hat k',0)$ and Taylor expand the term in parentheses about $\hat k,\hat k'=\hat k_{opt}$, we obtain
\be
\label{eq:gamma}
\Gamma^{(ind)}(\hat k,\hat k')\approx-\frac{\alpha^2}{4v_{\hat k_{opt}}} \frac{1-k_F^{-2}\left(\frac{1}{2}\left[\hat k+\hat k'\right]\right)^2}{\left[\kappa^2+(\hat k-\hat k')^2\right]^{1-\eta/2}}
\ee
$\hat k$ and $\hat k'$ are now measured relative to $\hat k_{opt}$. The momentum dependence in the numerator is schematic: $f(\hat k_{opt})$ is defined to equal $1$, but we don't pretend to get the various numbers right in the coefficients of the quadratic terms, which will not matter to the analysis.  For a highly anisotropic system the expression above follows after suitable rescaling of the components of $\hat k$ and $\hat k'$ to remove this anisotropy for for $|\hat k|,
|\hat k'|\ll k_F$ . As previously stated, $\phi(\hat k)$ will be peaked with width $\tilde \kappa$, so we write 
\be
\phi(\hat k)=\left[\tilde \kappa^{\frac{d-1}{2}} F\left(\frac{\hat k}{\tilde \kappa}\right)\right]^{-1}
\ee
Where by assumption
\be
\int d^{d-1}\hat k|\phi(\hat k)|^2=\int \frac{d^{d-1}\hat x}{|F(\hat x)|^2}=1
\ee
(All momentum integrals in this supplementary material are over the Fermi surface). $F$ can be unproblematically assumed to be real, and must be a suitably rapidly increasing function of its argument for large values of that argument. We need not assume a specific functional form for the eigenfunction, but only that it is characterized by a single momentum scale $\tilde \kappa \ll k_F$.   Since $\phi(\hat k)$ has the largest eigenvalue by assumption, the asymptotic dependence of $\tilde \kappa$ on $\kappa$ can be found by maximizing
\begin{align}
\lambda(\tilde \kappa,\kappa)\equiv &-\langle \phi |\Gamma^{(ind)}|\phi\rangle\\
=&-\int d^{d-1}\hat k d^{d-1} \hat k' \Gamma^{(ind)}(\hat k,\hat k') \phi(\hat k)\phi(\hat k')  \nn
\end{align}
with respect to $\tilde \kappa$. Introduce rescaled and rotated coordinates
\be
\hat k=\tilde\kappa \left(\vec K+\frac{1}{2}\vec Q\right)\qquad \hat k'= \tilde\kappa \left(\vec K-\frac{1}{2}\vec Q\right)
\ee
\begin{align}
\lambda(\tilde \kappa,\kappa)\propto &\tilde\kappa^{d-1}\int \left(1-\frac{\tilde\kappa^2}{k_F^2} |\vec K|^2\right)d^{d-1}\vec K G\left(\vec K,\vec Q\right) d^{d-1}\vec Q
\end{align}
Where the function $G$ is what's inside the $\vec Q$ integral:
\begin{align}
\label{eq:Qint}
&\int d^{d-1}\vec Q G(\vec K, \vec Q) \\
&=\int \frac{d^{d-1}\vec Q }{\left[\kappa^2+\tilde \kappa^2 |\vec Q|^2\right]^{1-\eta/2}}\cdot\frac{1}{F\left(\vec K+\frac{\vec Q}{2}\right)F\left(\vec K-\frac{\vec Q}{2}\right)}\nn 
\end{align}
The leading contribution to this integral in the limit $\kappa/\tilde\kappa \ll 1$ can be found by replacing $\vec Q$ with zero in the arguments of $F$, and subleading contributions can be found by suitable series expansions of $F$. These subleading contributions involve numbers of order one related to derivatives of $F$, but the actual numbers are not important for our asymptotic analysis. Once the leading contributions to the $\vec Q$ integral have been computed, the $\vec K$ integral just multiplies the result by $a-b\tilde \kappa^2/k_F^2$, where $a$ and $b$ are numbers of order one which don't affect asymptotic parameter dependence and which we thus ignore.
  
The form of the result depends on the quantity $d-3+\eta$. For the Ising nematic QCP in $d<3$ (including the formal case $d=3-\epsilon$) this quantity is negative, whereas it is zero in $d=3$, so we treat only these two cases. In both cases we ignore numbers of order one which will not affect the form of asymptotic results: 
\begin{itemize}
\item{{\bf Case 1}: $\mathbf {d-3+\eta<0}$}
\begin{align}
\lambda(\tilde \kappa,\kappa)\propto& \left(\kappa^{d-3+\eta} -\tilde\kappa^{d-3+\eta}\right)\left(1-\frac{\tilde\kappa^2}{k_F^2}\right)\\
\approx & \mbox{ const.} -\tilde\kappa^{d-3+\eta}-\kappa^{d-3+\eta}\frac{\tilde\kappa^2}{k_F^2}\nn
\end{align}
which is maximized for 
\begin{align}
\tilde \kappa \sim k_F \left(\frac{\kappa}{k_F}\right)^w, \mbox{ with}\\
w=\frac{3-d-\eta}{5-d-\eta}\nn
\end{align}
Since $0<w<1$ the width of $\phi(\hat k)$ clearly satisfies $\kappa \ll \tilde \kappa \ll k_F$. The eigenvalue is
\be
\lambda^{(ind)} \sim \alpha^2 \left(\frac{k_F}{\kappa}\right)^{3-d-\eta}
\ee
plus subleading corrections
\item{{\bf Case 2}: $\mathbf {d-3+\eta=0}$}

In this marginal case, the integral over the momentum transfer must be cut off (i.e. in Eq. \eqref{eq:Qint}, $Q$  must be limited above by $\sim k_F/\tilde\kappa$), but the analysis is still straightforward, with the result 
\be
\lambda(\tilde \kappa,\kappa)\propto \log\left[\frac{\tilde \kappa}{\kappa}\right]\left(1-\frac{\tilde\kappa^2}{k_F^2}\right)
\ee
which is maximized for
\be
\tilde\kappa\sim \frac{k_F}{\sqrt{\log\left[\frac{k_F}{\kappa}\right] }}
\ee
The expression for $\tilde \kappa$ in this case is valid up to corrections of order $\log (\log  (k_F/\kappa))$.  While $\sqrt{\log }$ is a very slowly increasing function of its argument, we still satisfy the condition $\kappa\ll \tilde \kappa \ll k_F$ when $\kappa / k_F\ll 1$. The eigenvalue is 
\be
\lambda^{(ind)} \sim \alpha^2 \log \left[\frac{k_F}{\kappa}\right]
\ee
plus subleading corrections
\end{itemize}

The approach above made only modest assumptions about the form of $\phi(\hat k)$, most notably that it is peaked at the minimum of $\Gamma^{(ind)}(\hat k, \hat k)$ and everywhere positive, considerations we know on general grounds will yield the ``ground state" of this operator.  A similar analysis could be applied to the ``first excited state(s)", which will  have a node at $\hat k_{opt}$ but otherwise have similar peak structure to $\phi(\hat k)$, and so on for the entire ladder of states which are localized about $\hat k=\hat k_{opt}$. The characteristic width of all these states is of order $\tilde \kappa$, and the fractional eigenvalue splittings between them of order $(\tilde\kappa/k_F)^2$.  There are a large but finite number of such states, of order $(k_F/\tilde\kappa)^{d-1}$.
\section{Eigenvalue splittings}
\label{sec:split}

  \begin{figure}%
\centering
\includegraphics[width=7cm]{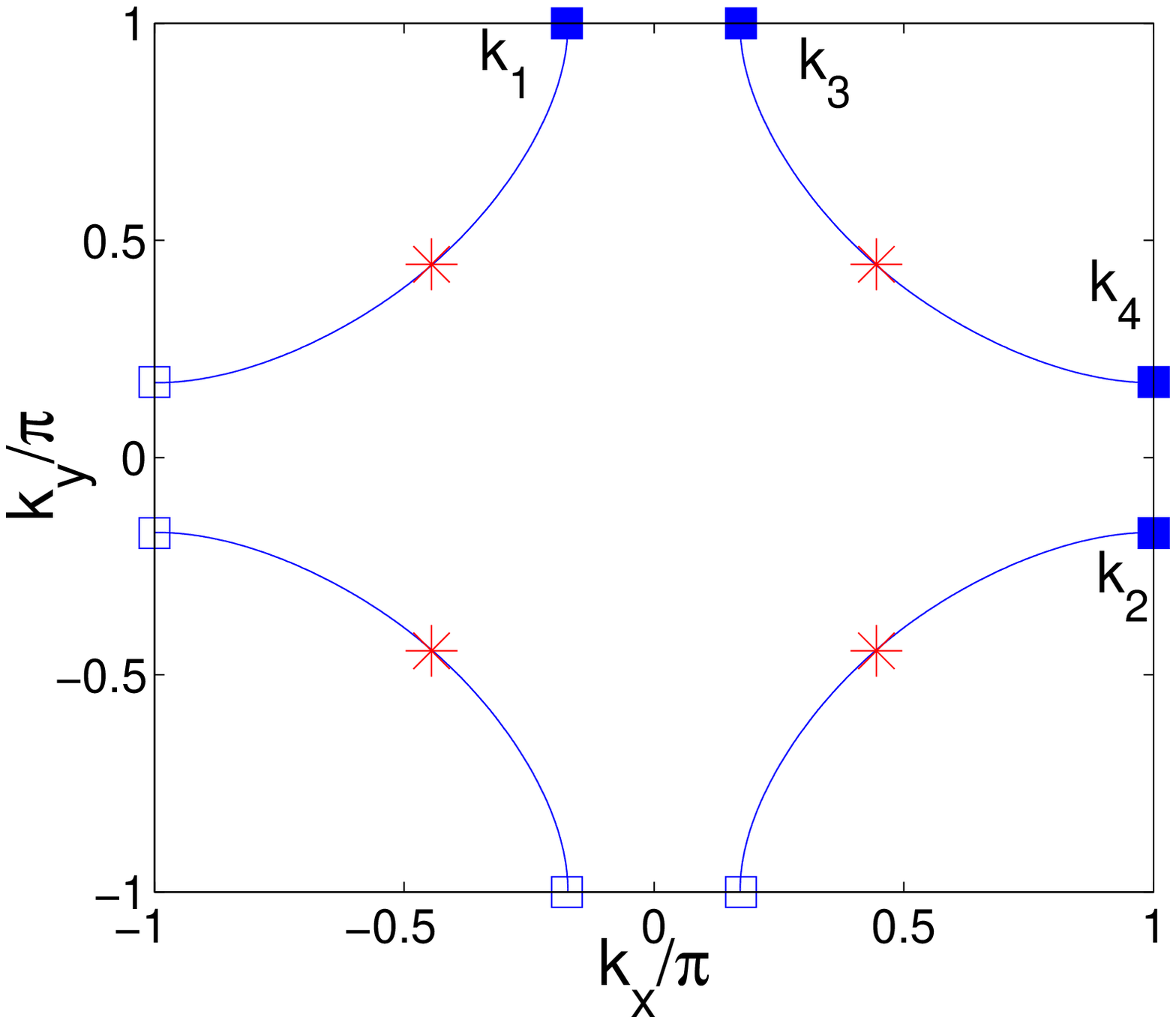}%
\caption{Two-dimensional cut of the Fermi surface considered in section \ref{sec:split}, labeling the points $\hat k_i$ where $\Gamma^{(ind)}(\hat k,\hat k)$ is strongest}
\label{fig:fermi}%
\end{figure}

The calculation of section \ref{sec:width} shows that, at each maximal point $\hat k_{opt}$ of $\Gamma^{(ind)}(\hat k, \hat k)$ on the Fermi surface, there is a ladder of eigenfunctions with similar eigenvalues.  When $d-3+\eta<0$, these eigenfunctions share a leading power-law divergence of their eigenvalue:
\be
\lambda^{(ind)} \sim \alpha^2\kappa^{d-3+\eta}
\ee
The splittings are small compared to $\lambda^{(ind)}$, but diverge as $\kappa$ vanishes:
\begin{align}
\delta \lambda \sim & \left(\frac{\tilde \kappa}{k_F}\right)^2 \lambda^{(ind)} \sim \alpha^2 \kappa^{2w+d-3+\eta}\\
\sim &\alpha^2\kappa^{\frac{-(d-3+\eta)^2}{5-d+\eta}} \nn
\end{align}
In the case $d-3+\eta =0$, the eigenvalue diverges logarithmically and the splittings within the ``ladder" are ${\cal O}(\alpha^2)$.

Looking at individual peaks of the gap function tells us nothing about the symmetry (s wave, p wave, or d wave).  We now quantify the small splittings among the different symmetry channels at the same rung of the ladder.  Let $|1\rangle$,$|2\rangle$,$|3\rangle$,$|4\rangle$ represent peaks (of the same rung) centered at the four maximal points $\hat k_{1,2,3,4}$ on the Fermi surface. The points $\hat k_i$ are related to each other by various reflections, as shown in Fig. 1.  The matrix $\Gamma^{(ind)}$ has four-fold rotation symmetry, so the eigenstates within this subspace are:
\begin{align}
|s\rangle\equiv& |1\rangle+|2\rangle+|3\rangle+|4\rangle\\
|d\rangle\equiv& |1\rangle-|2\rangle+|3\rangle-|4\rangle\nn\\
|p_x\rangle\equiv& |1\rangle-|3\rangle \nn \\
|p_y\rangle\equiv& |2\rangle-|4\rangle \nn
\end{align}
Let $\Gamma\equiv -\Gamma^{(ind)}$ in this section to save space. The eigenvalues are
\begin{align}
\lambda_s\equiv&\frac{\langle s|\Gamma|s\rangle}{\langle s|s\rangle} = \frac{4\lambda^{(ind)}+8\langle 1|\Gamma |2\rangle+4\langle 1|\Gamma |3\rangle}{4+8\langle 1|2\rangle+4\langle 1|3\rangle} \\\approx & \lambda^{(ind)}+2\langle 1|(\Gamma-\lambda^{(ind)})|2\rangle+\langle 1|(\Gamma-\lambda^{(ind)})|3\rangle \nn
\end{align}
\begin{align}
\lambda_d\equiv&\frac{\langle d|\Gamma|d\rangle}{\langle d|d\rangle} = \frac{4\lambda^{(ind)}-8\langle 1|\Gamma |2\rangle+4\langle 1|\Gamma |3\rangle}{4-8\langle 1|2\rangle+4\langle 1|3\rangle}\nn\\
\approx & \lambda^{(ind)}-2\langle 1|(\Gamma-\lambda^{(ind)})|2\rangle+\langle 1|(\Gamma-\lambda^{(ind)})|3\rangle \nn
\end{align}
\begin{align}
\lambda_p\equiv&\frac{\langle p_x|\Gamma|p_x\rangle}{\langle p_x|p_x\rangle}=\frac{\langle p_y|\Gamma|p_y\rangle}{\langle p_y|p_y\rangle} = \frac{2\lambda^{(ind)}-2\langle 1|\Gamma |3\rangle}{2-2\langle 1|3\rangle}\nn\\
\approx & \lambda^{(ind)}-\langle 1|(\Gamma-\lambda^{(ind)})|3\rangle\nn
\end{align}
where we have used $\langle i| \Gamma |i\rangle = \lambda^{(ind)}$. The splittings between the eigenvalues of states with different symmetry are dictated by matrix elements of $\Gamma-\lambda^{(ind)}$ between states with different peak positions. $\Gamma-\lambda^{(ind)}$ is just the tails of $\Gamma$, an object with only weak momentum dependence, so its matrix elements between different peaks will be dominated by large momentum transfer. If we take $i\neq j$ then
\begin{align}
\langle i|(\Gamma -\lambda^{(ind)})|j\rangle=&\int d^{d-1}\hat k d^{d-1}\hat k' \langle i|\hat k\rangle\langle \hat k|(\Gamma-\lambda^{(ind)})|\hat k'\rangle \langle \hat k' |j\rangle\nn\\
\approx & \Gamma(\hat k_i,\hat k_j) \int d^{d-1}\hat k d^{d-1} \hat k' \langle i|\hat k\rangle \langle \hat k' |j\rangle\nn\\
\propto& \Gamma(\hat k_i,\hat k_j) \tilde\kappa^{d-1}\nn\\
\propto & \alpha^2f\left(\hat k_i-\hat k_j,\frac{1}{2}(\hat k_i+\hat k_j)\right)^2\tilde\kappa^{d-1} 
\end{align}
At this level of approximation we find that $\langle 1|(\Gamma-\lambda^{(ind)})|3\rangle\sim \alpha^2\tilde\kappa^{d-1}$ but that $\langle 1|(\Gamma-\lambda^{(ind)})|2\rangle$ vanishes by symmetry.  To evaluate $\langle 1|(\Gamma-\lambda^{(ind)})|2\rangle$ we need to include the leading momentum dependence of $\Gamma(\hat k,\hat k')$ for $\hat k\approx \hat k_1$ and $\hat k'\approx \hat k_2$, finding $\langle 1|(\Gamma-\lambda^{(ind)})|2\rangle\sim \alpha^2\tilde\kappa^{d+1}$.  Our results for the splittings are:
\begin{align}
\lambda_s-\lambda_d\sim & \alpha^2 \tilde\kappa^{d+1}\\
\lambda_s-\lambda_p\approx \lambda_d-\lambda_p\sim &\alpha^2 \tilde\kappa^{d-1} \nn
\end{align}
The leading eigenstates of $\Gamma^{(ind)}$ are always $s$-wave, followed by $d_{x^2-y^2}$ and then degenerate $p_x$ and $p_y$ states (all of them with pronounced anisotropy, i.e. peaked with width $\tilde \kappa \ll k_F$ about the points $\hat k_{opt}$).  However, the differences between the eigenvalues of these states are due to large momentum transfer interactions and {\it vanish} as power laws as $\kappa \rightarrow 0$. By contrast, the eigenvalues themselves are controlled by critical small momentum transfer interactions and {\it diverge} as power laws. If we introduce a non-critical interaction $\Gamma^*$, its effect on the eigenvalue of some state $|x\rangle$ can be computed in first order perturbation theory.
\begin{align}
\delta\lambda^*=&\int d^{d-1}\hat k d^{d-1}\hat k' \langle x|\hat k\rangle\langle \hat k |\Gamma^*|\hat k'\rangle\langle \hat k' | x\rangle\\
\sim& \lambda^* \tilde\kappa^{d-1}
\end{align}
In the final relation we have assumed that $\Gamma^*$ has modest momentum dependence, and leading eigenvalue $\lambda^*$, whereas the state $|x\rangle$ is strongly peaked with width $\sim \tilde\kappa$.  Accordingly, if $\lambda^*\gg \alpha^2$ (as required for consistency far from criticality), $\Gamma^*$ will \textbf{always} determine the pairing symmetry.

\section{Crossover to ``strong enhancement" regime}
Far from criticality, where $|\delta x| = {\cal O}(1) $, $\lambda^{(ind)}\sim \alpha^2$. By assumption $\alpha^2 \ll \lambda^{(0)} < \lambda^*$, so this is in the regime of ``weak enhancement".  As we approach criticality, both $\lambda^{(ind)}$ and $\lambda^*$ grow, and we may or may not cross over into a regime in which $\lambda^{(ind)}\gg \lambda^*$, which we term ``strong enhancement".  In this section we clarify the conditions under which this crossover occurs in our regime of theoretical control, $|\delta x|\gg \alpha^{2/\gamma}$. The analysis once again depends on $d-3+\eta$:
\begin{itemize}
\item{{\bf Case 1}: $\mathbf {d-3+\eta<0}$}

In this case $\lambda^{(ind)}$ diverges as a power law on approach to criticality, i.e.
\be
\lambda^{(ind)}\sim\alpha^2 \left(\frac{\kappa}{k_F}\right)^{d-3+\eta}
\ee
By contrast, the growth of $\lambda^*$ is very slow except exponentially close to criticality:
\be
\lambda^*= \frac{\lambda^{(0)}}{1-\lambda^{(0)}\log \left[\frac{W}{\Omega}\right]}\approx \frac{\lambda^{(0)}}{1-\lambda^{(0)}\log \left[\frac{k_F}{\kappa}\right]}
\ee
Accordingly, the crossover will take place around the scale where $\lambda^{(ind)}=\lambda^{(0)}$, i.e.
\be
\frac{\kappa}{k_F}\sim\left(\frac{\alpha^2}{\lambda^{(0)}}\right)^{\frac{1}{3-d-\eta}} 
\ee
or equivalently
\be
 |\delta x|\sim \left(\frac{\alpha^2}{\lambda^{(0)}}\right)^{\frac{1}{\nu(3-d-\eta)}}\sim\left(\frac{\alpha^2}{\lambda^{(0)}}\right)^{\frac{1}{\gamma -(d-1)\nu}}
\ee

For the crossover to strong enhancement to occur within our regime of control $|\delta x|\gg\alpha^{2/\gamma}$, $\lambda^{(0)}$ must satisfy
\be
\lambda^{(0)} \ll \alpha^{2(d-1)\nu}
\ee

\item{{\bf Case 2}: $\mathbf {d-3+\eta=0}$}

In this case $\lambda^{(ind)}$ only diverges logarithmically:
\be
\lambda^{(ind)} \sim \alpha^2 \log \left[\frac{k_F}{\kappa}\right]
\ee
If we set the above expression equal to $\lambda^*$, the solution furthest from criticality is
\be
\log \left[\frac{k_F}{\kappa}\right] = \frac{1}{2\lambda^{(0)}}\left[1-\sqrt{1-\frac{4(\lambda^{(0)})^2}{\alpha^2}}\right]
\ee
The solution only exists for $\lambda^{(0)}<\alpha/2$. In this case we can say that $\lambda^{(0)} \sim \alpha ^{2-x}$, where $0<x\leq 1$.  The result is:
\begin{align}
\frac{\kappa}{k_F} \sim &\exp\left(-\frac{\#}{ \alpha^{x}}\right)\\
|\delta x| \sim & \exp\left(-\frac{\#}{\nu\alpha^x}\right)\ll \alpha^{\frac{2}{\gamma}}\nn
\end{align}
\end{itemize}
where $\#$ is a number of order one. When $d-3+\eta=0$, as is the case for the Ising QCP in $d=3$, the crossover to ``strong enhancement" occurs, if at all, exponentially close to criticality, outside the domain where our approach is controlled.  

In a quasi-2D system, there is a dimensional crossover from 2D to 3D at a value of $\kappa$ determined by the anisotropy of the bosonic theory in the UV.  A strong enhancement regime in $d=3$ may occur in our regime of control if this dimensional crossover occurs sufficiently close to the critical point. In particular, the crossover to strong enhancement could takes place when the bosons are effectively 2D, and only then cross over into 3D behavior, where the eigenvalue would be strongly enhanced but grow only logarithmically on approach to criticality.

\section{Numerical diagonalization of $\Gamma$}
Given a Fermi surface and a form of the pairing vertex $\Gamma(\hat k, \hat k')$, it is straightforward to find the eigenvalues and eigenstates numerically.  This is accomplished by discretizing the Fermi surface into $N$ equal patches, and diagonalizing the $N\times N$ matrix ${\underline \Gamma}$ defined by 
\be
{\underline \Gamma}^{n,m} = \Gamma(\hat k_n,\hat k_m)\cdot \frac{A}{N}
\ee
Here $\hat k_n$ is the center of the $n$th patch, and $A\sim k_F^{d-1}$ is the Fermi surface area (or length in $d=2$). In order to resolve features of the critical interaction $\Gamma^{(ind)}$, we must take $N$ large enough that $A/N \ll \kappa^{d-1}$.  

The asymptotic statements of sections \ref{sec:width} and \ref{sec:split} have been confirmed by this numerical approach for $d=2$, using values of $k_F/\kappa$ as high as $500$ and values of $N$ as high as $4000$.  In $d=3$ the calculation is much more time consuming for a given value of $\kappa$ due to the larger number of patches necessary. In addition, the asymptotic behaviors are (putatively) logarithmic rather than power law, requiring much smaller values of $\kappa$ to discern with certainty. The limited numerical diagonalization we have carried out in $d=3$ yields results consistent with out asymptotic analysis.

When we include both the boson-mediated interactions and the non-critical interactions, we have 
\be
\Gamma(\hat k,\hat k')=\Gamma^*(\hat k,\hat k') +\Gamma^{(ind)}(\hat k, \hat k')
\ee
For the numerical calculations presented in the paper we have assumed $\Gamma^*$ to be a separable d-wave interaction, i.e
\be
\Gamma^*(\hat k,\hat k') \propto \lambda^*\left(\cos \hat k_x-\cos \hat k_y\right)\left(\cos \hat k'_x-\cos \hat k'_y\right)
\ee
With this simplifying assumption the solutions of the eigenvalue equation depend on only two parameters: $\alpha^2/\lambda^*$ and $ \kappa/ k_F$.  Figure 1b of the paper shows results for $\lambda^*=10\alpha^2$, $\kappa / k_F = 0.1,0.01$, with $N=1000$ (larger values of $N$ yield identical results for these parameter choices).

\bibliographystyle{apsrev}
\bibliography{supplementary}